\documentclass[aps,prd,amsmath,amsfonts,amssymb,eqsecnum,nofootinbib,
  floatfix,secnumarabic,preprintnumbers,superscriptaddress,nobalancelastpage,
  onecolumn,notitlepage]{revtex4}
\usepackage{color,graphicx,hyperref,dsfont,slashed,multirow}
\usepackage{placeins}

\hypersetup{bookmarks=true,unicode=true,pdftoolbar=true,pdfmenubar=true,
  pdffitwindow=false,pdfstartview={FitH},pdftitle={mBLRISS},
  pdfauthor={~M.~Frank,\"{O}.~\"{O}zdal}, pdfsubject={Subject},
  pdfcreator={Creator},pdfproducer={Producer}, pdfkeywords={Non-minimal 
  supersymmetry}{Beyond the Standard Model Physics}{sneutrino dark matter},
  pdfnewwindow=true,colorlinks=true,
  linkcolor=blue,citecolor=magenta,filecolor=magenta,urlcolor=cyan}
\newcommand{\be}{\begin{equation}}
\newcommand{\ee}{\end{equation}}
\def\bsp#1\esp{\begin{split}#1\end{split}}
\renewcommand{\figureautorefname}{Fig.}

\def\singR{\mathcal{X}_{R}}
\def\singRbar{\mathcal{\overline{X}}_{R}}

\def\sectionautorefname~#1\null{Sec.~(#1)\null}
\def\subsectionautorefname~#1\null{sub--Sec.~(#1)\null}
\def\figureautorefname~#1\null{Fig.~#1\null}
\def\tableautorefname~#1\null{Table~#1\null}
\def\equationautorefname~#1\null{Eq.~#1\null}

\begin{document}
\preprint{CUMQ/HEP 195}

{\title{Exploring the  supersymmetric U(1)$_{B-L} \times$ U(1)$_{R}$ model with dark matter, muon $g-2$ and $Z^\prime$ mass limits}

\author{Mariana Frank}
\email{mariana.frank@concordia.ca}
\affiliation{Department of Physics, Concordia University 7141 Sherbrooke St. West, Montreal, QC,
	CANADA H4B 1R6}

\author{\"{O}zer \"{O}zdal}
\email{ozer.ozdal@concordia.ca}
\affiliation{Department of Physics, Concordia University 7141 Sherbrooke St. West, Montreal, QC,
 CANADA H4B 1R6}

\vspace{10pt}
\begin{abstract}
 
We study the low scale predictions of supersymmetric standard model extended by $U(1)_{B-L}\times U(1)_{R}$ symmetry, obtained from $SO(10)$ breaking via a left-right supersymmetric model, imposing universal boundary conditions. Two singlet Higgs fields  are responsible for the radiative $U(1)_{B-L}\times U(1)_{R}$ symmetry breaking, and a singlet fermion $S$ is introduced to generate neutrino masses through inverse seesaw mechanism. The lightest neutralino or sneutrino emerge as dark matter  candidates, with different  low scale implications. We find that the composition of the  neutralino LSP changes considerably depending on the neutralino LSP mass, from roughly half $U(1)_R$ bino,  half  MSSM bino, to singlet higgsino, or completely dominated by MSSM higgsino. The sneutrino LSP is statistically much less likely, and when it occurs it is a 50-50 mixture of right-handed sneutrino and the scalar $\tilde S$. Most of the solutions consistent with the relic density constraint
survive the XENON 1T exclusion curve for both LSP cases.  We compare the two scenarios and investigate parameter space points and find consistency with the muon anomalous magnetic moment  only at the edge of  $2\sigma$ deviation from the measured value. However, we find that the sneutrino LSP solutions could be ruled out completely by strict reinforcement of the recent $Z^\prime $ mass bounds. We finally discuss collider prospects for testing the model.

\end{abstract}


\maketitle

\section{Introduction}\label{sec:intro}
The discovery of the SM-like Higgs boson and nothing else presents a serious challenge to particle phenomenology. On one hand, the SM is incomplete, as it fails to explain properties such as the hierarchy problem, neutrino masses, cosmological inflation and dark matter. On the other hand, a Higgs mass of $125$ GeV presents a problem for the Standard Model (SM) ({\it e.g.}, electroweak vacuum instability), and for most of its extensions. So far, no clear directions for theoretical explorations or experimental solutions are indicated. Constructing and studying models which attempt to solve some of the outstanding problems in SM emerges as a viable alternative. Out of these, supersymmetry presents a partial solution to the hierarchy problem and  a clear one for dark matter. However, in its minimal incarnation, the minimal supersymmetric model (MSSM) requires squarks and gluinos in the multi-TeV range to explain such a low Higgs mass, raising a serious challenge for the LHC to find any signals.  

This issue may be resolved in models with extended gauge groups. In these models, additional $D$-term contributions to the Higgs mass matrices weaken considerably MSSM mass limits \cite{Haber:1986gz,Cvetic:1997ky,Ma:2011ea}. Depending on the models studied, these models can also resolve additional problems of MSSM. For instance, models with left-right symmetry \cite{Mohapatra:1995xd} can yield neutrino masses via the seesaw mechanism \cite{Mohapatra:1980yp, Schechter:1980gr, Schechter:1981cv}.
     
In \cite{Malinsky:2005bi}, an extended supersymmetric model based on $SU(3)_c \times SU(2)_L \times U(1)_R \times U(1)_{B-L} $ was proposed. The model can be embedded in $SO(10) $ SUSY-GUT, much like the left-right supersymmetric model, and generate a new seesaw mechanism for neutrino masses. The factor $U(1)_R$ can be thought off as remnant of a more complete $SU(2)_R$. Unlike the left-right supersymmetric model, which requires  Higgs triplet representations with vacuum expectation values (VEV) $v_R \sim 10^{15}$ GeV for obtaining neutrino masses and gauge unification, the symmetry in this model can be broken by singlet Higgs bosons (thought of as remnants of a doublet representation in left-right models), with VEVs in the TeV range, while still allowing for gauge coupling unification.  In \cite{Malinsky:2005bi}, the smallness of neutrino masses was explained as based on an inverse seesaw mechanism. The general features of the TeV scale soft-supersymmetry breaking parameters were  explored in \cite{DeRomeri:2011ie}, outlining conditions for models with intermediate scales obtained from breaking $SO(10)$.  The Higgs sector of the model was further explored, showing that a larger mass than that predicted by MSSM can be obtained. The parameter space was further explored in  \cite{Hirsch:2012kv}, where benchmarks, branching ratios, as well as lepton violation constraints were analyzed.

In this work, we concentrate on investigating, discriminating, and restricting the parameter space of the model using dark matter studies. We include up-to-date constraints on the spectrum coming from the Higgs signal strengths and mass data, and including LHC restrictions on squark and gluino masses, constraints on flavor parameters from the $B$ sector, as well as recent lower limits on the $Z^\prime$ mass.
Assuming universal scalar and gaugino masses, we show that the lightest supersymmetric particle (LSP) can be the sneutrino (which is different from the usual in this scenario, being a mixture of the right sneutrino and a gauge singlet fermion introduced to generate the inverse seesaw mechanism); or the lightest neutralino (which is favored to be a mixture of the two $U(1)$ binos). Relic density and indirect dark matter detection severely restrict the parameter space, as indeed does the recent limit on the $Z^\prime$ mass \cite{ATLAS:2017wce}. 
Within the parameter space allowed by dark matter limits, we analyze the consequences on sparticle spectra, the neutral Higgs sector and on the  anomalous magnetic moment of the muon, which shows more than $3~\sigma$ \cite{Bennett:2006fi} discrepancy with the SM prediction. Finally we investigate the possibilities of testing the model at the LHC. 

Our work is organized as follows. We provide a brief description of the model in Sec. \ref{sec:model}, capitalizing on more complete descriptions which have appeared previously. In Sec. \ref{sec:scan} we describe in detail the parameters of the model and constraints imposed on them. 
Dark matter phenomenology is explored in Sec. \ref{sec:DMpheno}, for both neutralino LSP \ref{subsec:neutralinoDM} and sneutrino LSP \ref{subsec:sneutrinoDM}. We then look at the consequences of our findings and compare the two scenarios in Sec. \ref{sec:comparisonLSP}, for the sparticle spectrum, the Higgs sector \ref{subsec:Higgssector} and the anomalous magnetic moment of the muon \ref{subsec:muong2}, and show that  imposing the $Z^\prime$ strict mass limitsbasically rules out  the sneutrino DM solutions in \ref{subsec:zprime}. We discuss possibilities for detection in Sec. \ref{sec:collider} and conclude in Sec. \ref{sec:conclusion}. We leave some relevant formulas for the Appendix.


\section{Model Description}\label{sec:model}

In this section, we describe the supersymmetric model under investigation briefly. This model, based on $SU(3)_c \times SU(2)_L \times U(1)_R \times U(1)_{B-L} $ (thereafter referred to as the BLRSSM) was first introduced in \cite{Malinsky:2005bi} and further studied in \cite{DeRomeri:2011ie, Hirsch:2011hg, Hirsch:2012kv}. The model emerges from breaking of supersymmetric $SO(10)$ to the SM through the following intermediary steps,
$$SO(10) \to SU(3)_C \times SU(2)_L \times SU(2)_R \times U(1)_{B-L} \to SU(3)_C \times SU(2)_L \times U(1)_R \times U(1)_{B-L} \to SU(3)_C \times SU(2)_L \times  U(1)_{Y} .$$
The advantages of this model are
\begin{itemize}
\item It is obtained by breaking of SO(10) through a left-right symmetric model, thus inheriting some of its attractive features \cite{Mohapatra:1996vg,Mohapatra:1995xd}; 
\item It is able to explain neutrino masses by the inverse seesaw  mechanism \cite{Malinsky:2005bi};
\item It preserves  gauge coupling unification of the MSSM, even when the breaking scale in the last step is of the order of the electroweak scale \cite{DeRomeri:2011ie};
\item It resolves the MSSM Higgs mass problem by yielding larger Higgs masses through additional $D$-terms in the soft-breaking potential, without resorting to heavy particles \cite{DeRomeri:2011ie};
\item It could yield signals differentiating it from MSSM, which may lie in different regions of SUSY parameter space;
\item It could provide different dark matter candidates and phenomenology, which in turn inform the study of direct and indirect searches.
\end{itemize}

The particle content of the model contains, in addition to the SM particles:
\begin{enumerate}
\item In the fermionic/matter sector,  an additional (right-handed) neutrino $N_i^c$, required for anomaly cancellation, and an additional singlet fermion $S$, needed for generating neutrino masses. Both these fermions come in 3 families and are accompanied by their scalar partners; 
\item In the bosonic/Higgs sector, two new Higgs fields, $\singR$ and $\singRbar$, remnants of $SU(2)_R$ doublets, needed to break $U(1)_R \times U(1)_{B-L} \to U(1)_Y$, and their fermionic partners;
\item In the gauge sector, an additional neutral gauge field, $Z^\prime$, which emerges from the mixing of the neutral gauge fields of  $SU(2)_L ,  U(1)_R $ and $ U(1)_{B-L} $, $(W^0, B_{R}, B_{B-L})$, and its fermionic partner.
\end{enumerate}
In a sense, the model described here is minimal: however it requires an extra ${\cal Z}_2$ matter parity to avoid breaking of $R$-parity  \cite{Hirsch:2012kv}.

The superpotential in this model is described by
\begin{eqnarray}
	W&=&\mu H_{u}H_{d}+Y_{u}^{ij}Q_{i}H_{u}u^{c}_{j}-Y_{d}^{ij}Q_{i}H_{d}d^{c}_{j}-Y_{e}^{ij}L_{i}H_{d}e^{c}_{j} \nonumber \\
&+&Y_{\nu}^{ij}L_{i}H_{u}N^{c}_{i}+ Y^{ij}_{s}N^{c}_{i}\singR S - \mu_{R}\singRbar \singR + \mu_{S}S S \, ,
\label{superpotential}
\end{eqnarray}
where the first line of Eq.(\ref{superpotential}) contains the usual terms of the MSSM, while the second line includes the additional interactions from the right-handed neutrino $N^{c}_{i}$ and the singlet Higgs fields $\singRbar$, $\singR$  with -1/2 and +1/2 $B - L$, and +1/2 and -1/2 $R$ charges, respectively. The first term of the second line in superpotential describes the Yukawa interactions between neutrinos, and $Y_{\nu}^{ij}$ is the Yukawa coupling associated with these interactions. In a similar manner, $Y^{ij}_{s}$ represents the Yukawa coupling among $N^{c}_{i}$, $\singR$ and $S$. Moreover, $\mu_{R}$ is  similar to $\mu'$ term of the $B-L$ extension of supersymmetric model (BLSSM) and stands for bilinear mixing between $\singR$ and $\singRbar$ fields. Note that there is also a $\mu_{S}$ term to generate non-zero neutrino masses with inverse seesaw mechanism, and as customary, it is restricted to have a  low value, as it cannot give important contributions to any other sector except for neutrinos. Contrary to BLSSM \cite{DelleRose:2017smp,Un:2016hji,Basso:2015xna}, where neutrinos have Majorana mass terms, $N^{c}_{i}$ fields interact with $\singR$ and $S$ through $ Y^{ij}_{s}N^{c}_{i}\singR S$ term, and lead to SM-singlet pseudo-Dirac mass eigenstates. Besides, the interaction of the $SU(2)_{L}$ singlet Higgs fields $\singR$, $S$ and $N^{c}_{i}$ yield a significant contribution to the masses of the extra Higgs bosons. Implementing the inverse seesaw mechanism into model allows $Y_{\nu}^{ij}$ and $Y^{ij}_{s}$ to be at the order of unity. Hence, the contribution from the right-handed neutrino sector to the Higgs boson cannot be neglected and yields a different low scale phenomenology from MSSM and BLSSM with inverse seesaw mechanism \cite{Abdallah:2017gde,Khalil:2015wua,Khalil:2015naa}. 

The soft-breaking Lagrangian terms in the model are
\begin{eqnarray}
	-{\cal L}_{SB,W}&=&- B_\mu (H_u^0 H_d^0- H^-_d H_u^+) - B_{\mu_R} \singR \singRbar + A_u ({\tilde u}^{\star}_{R, i} {\tilde u}_{L,j} H^0_u- {\tilde u}^{\star}_{R, i} {\tilde d}_{L,j} H^+_u) + A_d ({\tilde d}^{\star}_{R, i} {\tilde d}_{L,j} H^0_d-{\tilde d}^{\star}_{R, i} {\tilde u}_{L,j} H^-_d)\nonumber\\
	&+&A_e ({\tilde e}^{\star}_{R, i} {\tilde e}_{L,j} H^0_d- {\tilde e}^{\star}_{R, i} {\tilde \nu}_{L,j} H^-_d) +A_\nu ({\tilde \nu}^{\star}_{R, i} {\tilde \nu}_{L,j} H^0_u- {\tilde e}^{\star}_{R, i} {\tilde \nu}_{L,j} H^-_u) + A_{s, ij}\singR {\tilde \nu}_{R,i} {\tilde S}+ {\rm h.c.} \, ,\nonumber\\
	-{\cal L}_{SB,\phi}&=& m_{\singR}^2 | \singR |^2 +m_{\singRbar}^2 | \singRbar |^2 +m_{H_d}^2 ( | H_d^0 |^2 +|H_d^-|^2) +m_{H_u}^2 (| H_u^0 |^2 +|H_u^+|^2) + m^2_{q,ij}({\tilde d}^\star_{L,i} {\tilde d}_{L,j}+ {\tilde u}^\star_{L,i} {\tilde u}_{L,j}) \nonumber \\
	&+&m^2_{d,ij}{\tilde d}^\star_{R,i} {\tilde d}_{R,j} +m^2_{u,ij}{\tilde u}^\star_{R,i} {\tilde u}_{R,j} +m^2_{l,ij}({\tilde e}^\star_{L,i} {\tilde e}_{L,j}+{\tilde \nu}^\star_{L,i} {\tilde \nu}_{L,j}) + m^2_{e,ij}{\tilde e}^\star_{R,i} {\tilde e}_{R,j}+m^2_{\nu,ij}{\tilde \nu}^\star_{R,i} {\tilde \nu}_{R,j} +m^2_{s,ij}{\tilde S}^\star_{i} {\tilde S}_{j}\nonumber \\
	-{\cal L}_{SB, \lambda}&=&\frac12 \left(  M_1\lambda_B^2 +M_2 \lambda_W^2 +M_3 \lambda_g^2 + 2M_{B_R} \lambda_B \lambda_R +{\rm h.c.} \right)\, ,
\label{softbreaking}
\end{eqnarray}
which contain triple scalar interactions, scalar masses and masses for the gauginos of all gauge groups, denoted by $\lambda$'s. 

The $U(1)_R \times U(1)_{B-L}$ symmetry is broken spontaneously to $U(1)_Y$ by the vacuum expectation values (VEVs) of $\singR$ and $\singRbar$
\begin{equation}
\langle \singR \rangle =\frac{v_{\singR}}{\sqrt{2}}\, , \qquad \langle \singRbar \rangle =\frac{v_{\singRbar}}{\sqrt{2}}\, ,
\end{equation}
while $SU(2)_L \times U(1)_Y$ is broken further to $U(1)_{EM}$ by the VEVs of the Higgs doublets
\begin{equation}
\langle H_d^0 \rangle =\frac{v_d}{\sqrt{2}}\, , \qquad \langle H_u^0 \rangle =\frac{v_u}{\sqrt{2}}\, .
\end{equation}
We denote $v_R^2= v_{\singR}^2 + v_{\singRbar}^2$ and $\displaystyle \tan \beta_R= \frac{v_{\singR}}{v_{\singRbar}}$, in analogy with $v^2=v_d^2+v_u^2$, $\displaystyle \tan \beta=\frac{v_u}{v_d}$. 
The spectrum for this model, including particle masses, neutrino seesaw, mixing of gauge bosons and the neutralino sector has been discussed before \cite{Hirsch:2011hg}, and we do not repeat it here. In what follows we concentrate on scanning the model parameters first by imposing Higgs sector, particle masses and other low energy restrictions, and then looking for dark matter candidates and resolution of the anomalous magnetic moment of the muon, thus restricting the parameter space to region where these conditions are satisfied.


\section{Scanning Procedure and Experimental Constraints}
\label{sec:scan}

We proceed to analyze the model by scanning the fundamental parameter space of BLRSSM. We use  \textsc{SPheno} 3.3.3 package \cite{Porod:2003um,Porod:2011nf} obtained from the model implementation in  \textsc{Sarah} 4.6.0 \cite{Staub:2008uz,Staub:2010jh}. This package employs renormalization group equations (RGEs),  modified by the inverse seesaw mechanism  to evolve  Yukawa and gauge couplings from  $M_{{\rm GUT}}$ to the weak scale, where $M_{{\rm GUT}}$ is determined by the requirement of gauge coupling unification. We do not strictly enforce the solutions to unify at $M_{{\rm GUT}}$,  since a few percent deviation is allowed due to unknown GUT-scale threshold corrections \cite{Lucas:1995ic}. $M_{{\rm GUT}}$ is thus dynamically determined by the  requirement of gauge unification, that is  $ g_{L} = g_{R} = g_{B-L} \approx g_{3}$, with subindices denoting the gauge couplings associated with $ SU(2)_{L}, SU(2)_{R} ,U(1)_{B-L} $ and $SU(3)_{C}$ respectively. With boundary conditions determined at $M_{{\rm GUT}}$, all the soft supersymmetry breaking (SSB) parameters along with the gauge and Yukawa couplings are evolved to the weak scale.

\begin{table}
  \setlength\tabcolsep{7pt}
  \renewcommand{\arraystretch}{1.4}
  \begin{tabular}{c|c||c|c}
    Parameter      & Scanned range& Parameter      & Scanned range\\
    \hline
    $m_0$          & $[0., 3.]$~TeV   & $v_{R}$               & $[6.5, 20.]$~TeV\\
    $M_{1/2}$      & $[0., 3.]$~TeV   & $diag(Y_{\nu}^{ij})$  & $[0.001, 0.99]$\\
    $A_0/m_0$      & $[-3., 3.]$      & $diag(Y_{s}^{ij})$    & $[0.001, 0.99]$\\
    $\tan\beta$    & $[0., 60.]$      & {\rm sign of} $\mu$   & {\rm positive} \\
    $\tan\beta_R$  & $[1., 1.2]$    & {\rm sign of} $\mu_R$ & {\rm positive or negative} \\ 
  \end{tabular}
  \caption{\label{tab:scan_lim} Scanned parameter space.}
\end{table}

We performed random scans over the parameter space, as illustrated in \autoref{tab:scan_lim}, imposing  universal boundary conditions for scalar and gaugino masses. We comment briefly first on the parameters chosen, and then on the constraints included. Here $m_{0}$ corresponds the mass terms for all scalars, and $M_{1/2}$ represents the mass terms for all gauginos, including the ones associated with the $U(1)_{B-L}$ and $U(1)_{R}$ gauge groups. In setting the ranges for the free parameters, we scan scalar and gaugino SSB mass terms between 0--3 TeV,  regions which yield sparticle masses at the low scale, especially the LSP. 

Here $A_{0}$ is the  trilinear scalar interaction coupling coefficient, and we adjusted its range to avoid charge and/or color breaking minima, which translates into $\lvert A_{0} \rvert \lesssim 3 m_{0}$ \cite{Kusenko:1996vp,Chattopadhyay:2014gfa}. Also,  $\tan\beta$ is  the ratio of vacuum expectation values of the MSSM Higgs doublets $v_u/v_d$, while $\tan\beta_{R}$ which denotes the ratio of vacuum expectation values of $v_{\singR}/v_{\singRbar}$, is also free parameter in this model. Practically however, $\tan\beta_{R}$ is required to be close to 1, in order to prevent large $D$-term contributions to the sfermion masses and to avoid tachyonic solutions. The VEV $v_{R}$ represents the vacuum expectation value which breaks extra $U(1)_{B-L} \times U(1)_{R}$ symmetry. Since the breaking scale of the extra symmetry  plays a crucial role in determining the $Z^\prime$ mass,  the gauge boson associated with $U(1)_{B-L} \times U(1)_{R}$ symmetry, we scan $v_{R}$ between 6.5 and 20 TeV to obtain $Z^\prime$ boson masses consistent with the current experimental bounds. 

The parameter $\mu$ is the bilinear mixing of the MSSM doublet Higgs fields, while $\mu_{R}$ is the bilinear mixing  of the $SU(2)_{R}$ remnants Higgs fields, which are singlet under $SU(2)_{L}$ symmetry. The values of $\mu$ and $\mu_{R}$ can be determined by the radiative electroweak symmetry breaking (REWSB) but their signs cannot; thus, only their signs remain as free parameters. Since the model contributions to muon anomalous magnetic moment are related to the sign of $\mu M_{1/2}$, we scan over positive  $\mu$ values, but we accept both negative and positive solutions of $\mu_R$,  while requiring solutions consistent with experimental   predictions, and favoring solutions which improve upon the SM predictions for the muon $g-2$ factor.
The superpotential of the model also includes a $\mu_S$ parameter, which  yields non-zero neutrino masses via the inverse seesaw mechanism. However, $\mu_S$ is constrained to be small, so that it cannot effect any supersymmetric particle masses or decays. We also fixed the top quark mass to its central value ($m_t$ = 173.3 GeV) \cite{Group:2009ad} in our scan. The Higgs boson mass is very sensitive to the top quark mass,  and small changes in its value can shift Higgs boson mass by 1-2 GeV \cite{Gogoladze:2011aa,Ajaib:2013zha}, although it does not significantly affect sparticle masses \cite{Gogoladze:2011db}.  Hence, we scan both diag($Y_{\nu}^{ij}$) and diag($Y^{ij}_{s}$) between 0.001--0.99, though the inverse seesaw mechanism prefers values of order 1. 

\begin{table}{
		\setlength\tabcolsep{7pt}
		\renewcommand{\arraystretch}{1.6}
		\begin{tabular}{l|c|c||l|c|c}
			Observable & Constraints & Ref. & Observable & Constraints & Ref.\\
			\hline
			$m_{h_1} $ & $ [122,128] $ GeV                                & \cite{Chatrchyan:2012xdj} &
			$m_{\widetilde{t}_1} $                                 & $ \geqslant 730 $ GeV & \cite{Olive:2016xmw}\\
			$m_{\widetilde{g}} $                                     & $ > 1.75 $ TeV & \cite{Olive:2016xmw} &
			$ m_{\chi_1^\pm} $                                    & $ \geqslant 103.5 $ GeV & \cite{Olive:2016xmw} \\
			$m_{\widetilde{\tau}_1} $                                & $ \geqslant 105 $ GeV & \cite{Olive:2016xmw} & 
			$m_{\widetilde{b}_1} $                                 & $ \geqslant 222 $ GeV & \cite{Olive:2016xmw}\\
			$m_{\widetilde{q}} $                                     & $ \geqslant 1400 $ GeV & \cite{Olive:2016xmw} &
			$m_{\widetilde{\tau}_1} $                              & $ > 81 $ GeV & \cite{Olive:2016xmw} \\
			$m_{\widetilde{e}_1} $                                   & $ > 107 $ GeV & \cite{Olive:2016xmw} &
			$m_{\widetilde{\mu}_1} $                               & $ > 94 $ GeV & \cite{Olive:2016xmw} \\
			$\chi^2(\hat{\mu})$                                    & $\leq 3 $ & - &
			BR$(B^0_s \to \mu^+\mu^-) $ & $[1.1,6.4] \times 10^{-9}$  &
			\cite{Aaij:2012nna} \\ 
			$\displaystyle  \frac{{\rm BR}(B \to \tau\nu_\tau)}
			{{\rm BR}_{SM}(B \to \tau\nu_\tau)} $ & $ [0.15,2.41] $ &
			\cite{Asner:2010qj} &
			BR$(B^0 \to X_s \gamma) $ & $  [2.99,3.87]\times10^{-4} $ &
			\cite{Amhis:2012bh}\\
			$m_{Z^{\prime}} $                                        & $ > 3.5 $ TeV & \cite{ATLAS:2017wce} &
			$\Omega_{DM}h^{2} $                                            &  [0.09-0.14] & \cite{Komatsu:2010fb,Spergel:2006hy}                      \\  
		\end{tabular}
		\caption{\label{tab:constraints} Current experimental bounds imposed on the scan for consistent solutions.}}
	\end{table}

In scanning the parameter space, we use the interface which employs Metropolis-Hasting algorithm described in \cite{Belanger:2009ti}. All collected data points satisfy the requirement of REWSB. After collecting the data, we  impose current experimental mass bounds on all the sparticles and SM-like Higgs boson as highlighted in \autoref{tab:constraints}. Although we restrict the SM-like Higgs boson to lie between 122-128 GeV with 3 GeV uncertainty, we also employed \textsc{HiggsBounds} 4.3.1 package \cite{Bechtle:2013wla} to compare our Higgs sector predictions with the experimental cross section limits from the LHC, and  we require agreement with Higgs boson decay signal strengths at tree level,   $h \to WW^\star$, $h \to ZZ^\star$ and $h \to b \bar
{b}$. Thus using the mass-centered $\chi^2 $, and selecting the parametrization for the Higgs mass uncertainty as ``box" we employed \textsc{HiggsSignals} 1.4.0 package \cite{Bechtle:2013xfa} and bounded the solutions which yield total $\chi^2(\hat{\mu}) \leqslant $ 3. Another constraint comes from rare $B$-decay processes, $ B_s \rightarrow \mu^+ \mu^- $ \cite{Aaij:2012nna}, $b \rightarrow s \gamma$ \cite{Amhis:2012bh} and $B_u\rightarrow\tau \nu_{\tau}$ \cite{Asner:2010qj}. The $B$-meson decay into a muon pairs, in particular,  constrains the parameter space since there the SM predictions are consistent with the experimental measurements.  The supersymmetric contributions are proportional to $(\tan\beta)^6/m_{A_i}^{4}$ and constrained to be small. Hence, $m_{A_i}$ has to be heavy enough ($m_{A_i}\sim$ TeV) to suppress the supersymmetric contributions for large $\tan\beta$ values. In addition to these limitations, dark matter observations severely restrict the parameter space, requiring the  LSP to be stable and electric and color neutral, which excludes a significant portion of parameter space where stau is the LSP. We concentrate on two different data sets, one with the neutralino being the LSP, and one where sneutrino is the LSP, and we shall distinguish these two scenarios throughout our investigations. We employ \textsc{micrOMEGAs} 4.3.1 package \cite{Belanger:2014vza} and tag the solutions which yield consistent relic density within  20\% uncertainty range provided from WMAP data \cite{Komatsu:2010fb,Spergel:2006hy} as specified in \autoref{tab:constraints}. Apart from relic abundance constraint, we do not impose any restriction from the dark matter experiments. All the experimental restrictions mentioned above are listed in \autoref{tab:constraints}.

\section{Dark matter phenomenology}
\label{sec:DMpheno}

For either neutralino or sneutrino to be viable candidates for dark matter, they must yield the correct  level of relic abundance for thermal dark matter production in the early Universe, determined very precisely as the amount of non-baryonic dark matter in the energy-matter of the Universe, $\Omega_{DM}h^2=0.1199\pm 0.0027$ \cite{Ade:2013zuv}, with $\Omega_{DM}$ being the energy density of the dark matter with respect to the critical energy density of the universe, and $h$  the reduced Hubble parameter. 

In addition, as the lack of any dark matter signals in either direct or indirect dark matter detection experiments confront our theoretical expectations,  these must satisfy increasingly severe constraints from experiments. The interaction of dark matter with detector nuclear matter can be spin-dependent or spin-independent.   The spin-dependent scattering can only happen for odd-numbered nucleons in the nucleus
of the detector material, while in spin-independent (scalar) scattering, the coherent scattering of all
the nucleons in the nucleus with the DM are added in phase. Consequently, 
in direct detection experiments, the experimental sensitivity to spin-independent (SI) scattering is much 
larger than the sensitivity to spin-dependent scattering, and thus we shall concentrate on the former.  

We proceed as follows.  First, we analyze the consequences of having the lightest neutralino as the dark matter candidate. Using the results in the previous sections, we explore the parameter space of the model which is consistent with this assumption. We follow in the next subsection with the parameter restrictions for sneutrino dark matter.

\subsection{Neutralino Dark Matter}
\label{subsec:neutralinoDM}

In this subsection, we concentrate on analyzing the consequences on the mass spectrum of the BLRSSM obtained by scanning over the parameter space given in \autoref{tab:scan_lim} where lightest neutralino ($\widetilde{\chi}_1^0$) is always the LSP, and highlight the solutions compatible with the constraints showed in \autoref{tab:constraints}. We start with \autoref{fig:freeparams} which displays the allowed parameter regions, with plots in $ m_{0} - M_{1/2} $, $ m_0 - A_0/m_0 $  and $ M_{1/2} - \tan{\beta}$ planes. Throughout the graphs, all points satisfy REWSB. Blue points satisfy all experimental mass bounds, signal strengths of SM-like Higgs boson and rare $B$-decay constraints given in \autoref{tab:scan_lim}. Red points obey the above mentioned constraints, as well as  relic density bounds, 0.09 $ \leq \Omega_{DM}h^{2} \leq$ 0.14. The $ m_{0} - M_{1/2} $ plane shows that solutions for $M_{1/2} \lesssim $ 800 GeV are excluded by the constraints  in \autoref{tab:constraints}, and the requirement of  consistent relic density (red points) excludes a significant portion of the LHC allowed region (blue points). For $M_{1/2}\sim$ 1 TeV, $m_0$ is bounded between 2--3 TeV, and low $m_0$ values can survive for larger $M_{1/2}$. On the other hand, the $ m_0 - A_{0}/m_0 $ panel shows that the regions with larger $m_0$ values prefer positive values of the  trilinear scalar interaction strength $A_0$, while almost all solutions with consistent relic density  have positive $A_{0}$ parameter. Unlike the  $B-L$ Supersymmetric Standard Model (BLSSM) \cite{Un:2016hji}, where negative $A_{0}$ solutions for $m_0 \geq$  1 TeV do not satisfy REWSB, here all LSP constraints can be fulfilled for this portion of parameter space, while only the relic density constraint imposes positivity of $A_0$. The $ M_{1/2} -\tan{\beta}$ plot indicates that it is possible to find solutions with
0.09 $ \leq \Omega_{DM}h^{2} \leq$ 0.14 only for large $\tan \beta$ values, 40 $ \leq \tan\beta \leq$ 60,  although it is easier to satisfy LHC limitations for low $\tan\beta$ values.
\begin{figure}
	\centering
	\includegraphics[scale=0.31]{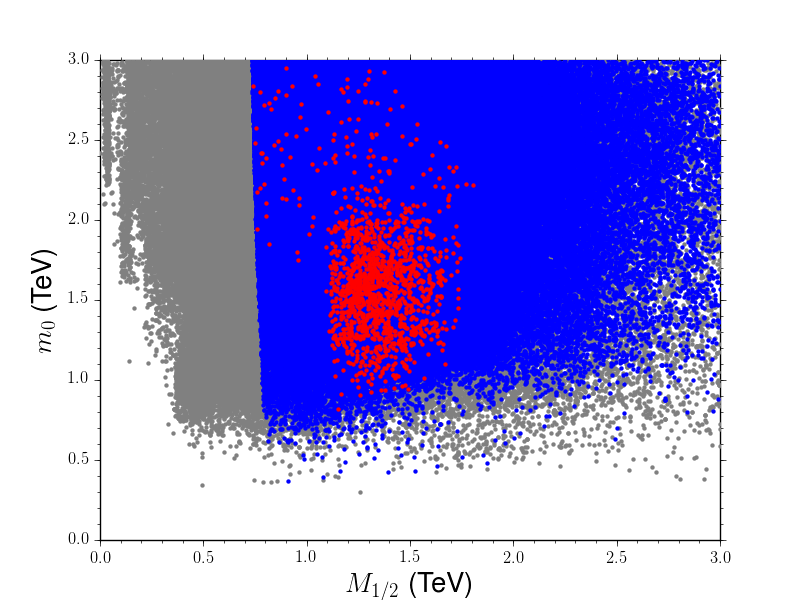}
	\includegraphics[scale=0.31]{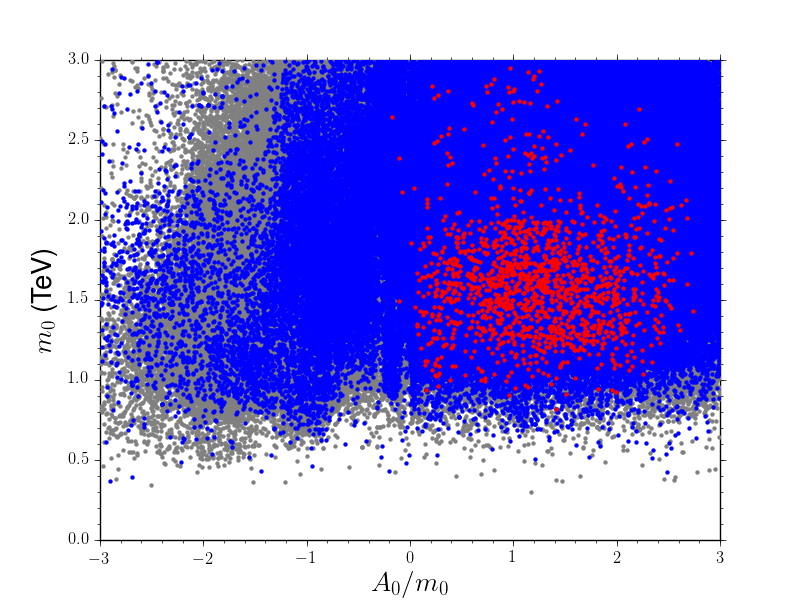}
	\includegraphics[scale=0.31]{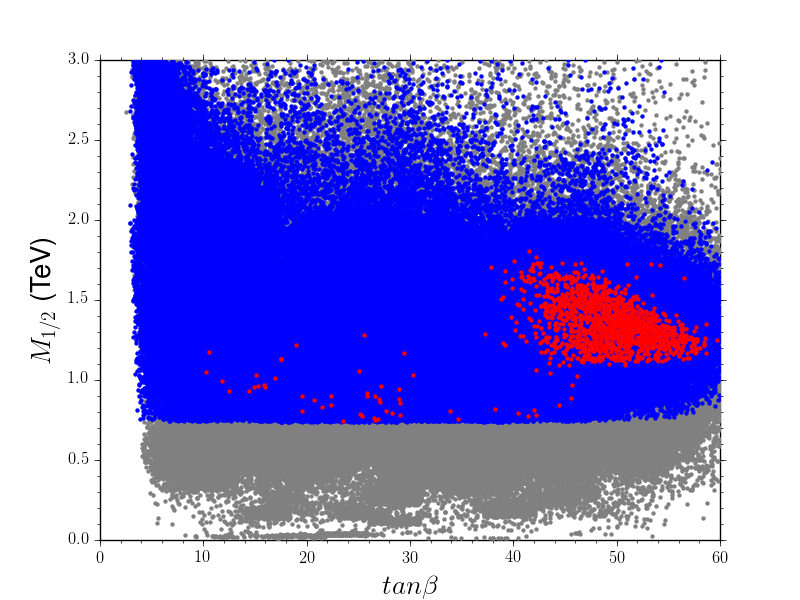}
	\caption{ Parameter scans for neutralino LSP scenario.  (Left) $ m_{0} $ vs $ M_{1/2} $, (center) $ m_{0} $ vs $A_{0}/m_{0}$ and (right) $ M_{1/2}$ vs $\tan{\beta}$. All points are consistent with REWSB and neutralino LSP. Blue points satisfy all the experimental limits listed in \autoref{tab:constraints}. Red points form a subset of blue, and represent solutions consistent with the relic density constraint.}
	\label{fig:freeparams}
\end{figure}

In \autoref{fig:sparticles}, we show specific results for the determination of sparticle mass spectrum, with plots in (top left) $ m_{\widetilde{t}_1} - m_{\widetilde{\chi}_1^0} $, (top right) $ m_{\widetilde{b}_1} - m_{\widetilde{\chi}_1^0} $, (bottom left) $ m_{\widetilde{\chi}_1^{\pm}} - m_{\widetilde{\chi}_1^0}$ and (bottom right) $ m_{\widetilde{\tau}_1} - m_{\widetilde{\chi}_1^0}$ planes. The color coding is the same as \autoref{fig:freeparams}. Furthermore, the mass region in which the two masses are degenerate is displayed as a solid green line. We note  that the LSP neutralino solutions consistent with the relic density bound can be obtained only when 300 GeV $ \leq m_{\widetilde{\chi}_1^0} \leq $ 800 GeV. As can be seen from $ m_{\widetilde{t}_1} - m_{\widetilde{\chi}_1^0} $ and $ m_{\widetilde{b}_1} - m_{\widetilde{\chi}_1^0} $ planes, we find that stop and sbottom masses have to  be at least $\sim$ 1.5 TeV and 2 TeV respectively to fulfill all the restrictions. Even though it is possible to find light stop solutions ($m_{\widetilde{t}_1} \leq $ 1 TeV) when 340 GeV $ \leq m_{\widetilde{\chi}_1^0} \leq $ 550 GeV, the relic density condition is not satisfied for these solutions. Moreover, unlike the results of BLSSM \cite{Un:2016hji} where the lightest chargino masses are always above 600 GeV, here the $ m_{\widetilde{\chi}_1^{\pm}} - m_{\widetilde{\chi}_1^0}$ plot shows that there is a region of parameter space where lightest chargino solutions is nearly degenerate with the lightest neutralino when 300 GeV $ \leq m_{\widetilde{\chi}_1^0} \leq $ 500 GeV. These solutions correspond to the case where the lightest chargino decays into the neutralino LSP and $W/W^\star$ boson ($\widetilde{\chi}_1^{\pm} \to \widetilde{\chi}_1^0 + W^\pm(W^{\star \pm})$), and the branching ratio for this channel is almost 1. On the bottom right panel, the $ m_{\widetilde{\tau}_1} - m_{\widetilde{\chi}_1^0}$ plane illustrates the stau mass along with the LSP neutralino mass. There is a parameter space around $m_{\tilde\chi^0} \sim 600 $ GeV, where stau mass is almost degenerate with the LSP neutralino and becomes the next to lightest supersymmetric particle (NLSP), but also for a region of the parameter space, the stau can be much heavier than the neutralino LSP.  The lightest stau NLSP solutions compatible with the relic density constraint occur around 500 GeV. One can choose one of these solutions and study relevant neutralino annihilation processes mediated by a light stau \cite{Calibbi:2013poa}.

The bottom  plots in \autoref{fig:sparticles},  show our results for the sparticle spectrum for the gluino and sneutrinos, with the plots in $ m_{\widetilde{q}} - m_{\widetilde{g}} $, (where $\widetilde{q}$ represents squarks from the first two families), and $ m_{\widetilde{\nu}_1} - m_{\widetilde{\chi}_1^0} $ planes. The $ m_{\widetilde{q}} - m_{\widetilde{g}} $ plane shows that squarks masses for the first two families and gluino masses should be heavier than 2 TeV but  lighter than 4 TeV (light blue points). Although the relic density condition and the current ATLAS experimental limit \cite{ATLAS:2017cjl} strictly constrain the crucial portion of the parameter space, most of the solutions are consistent with this experimental exclusion. Finally, the $ m_{\widetilde{\nu}_1} - m_{\widetilde{\chi}_1^0} $ plane reveals that it is hard to find solutions with sneutrino as the  supersymmetric NLSP  if we require consistency with the relic density bound, and the lightest sneutrino solutions  satisfying  all bounds can be obtained at around 1 TeV.  

Note that the graphs contain also information on the composition of the neutralino LSP. As can be seen from gluino vs squarks panel, light red points, which represent the mixed or higgsino-like neutralino LSP solutions consistent with the relic density bounds, are mostly found under the yellow  curve (the excluded region). Light blue points representing  mixtures of $R$-bino and $B-L$ bino (gauginos of $U(1)_R$ and $U(1)_{B-L}$, respectively) neutralino LSPs are mostly located within the 1 sigma error of the yellow line.
\begin{figure}
	\centering
	\includegraphics[scale=0.40]{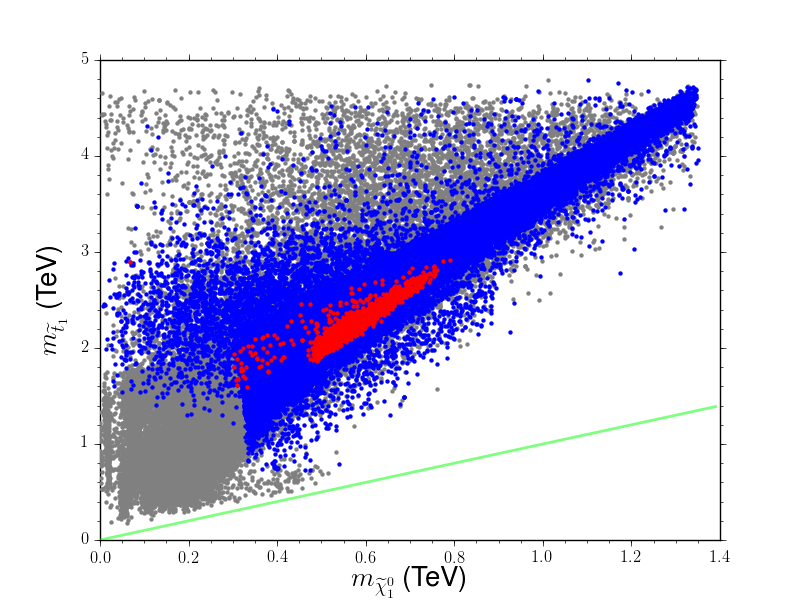}\hspace{0.5cm}
	\includegraphics[scale=0.40]{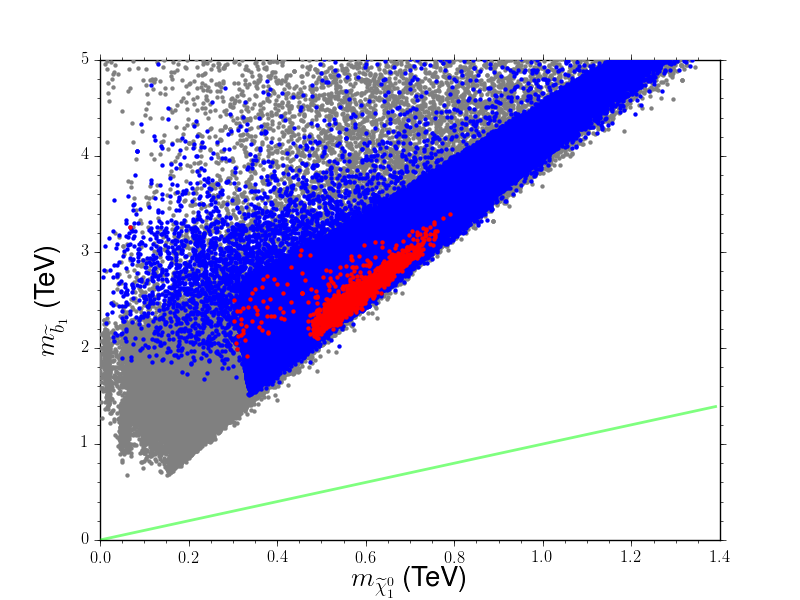}\\
	\includegraphics[scale=0.40]{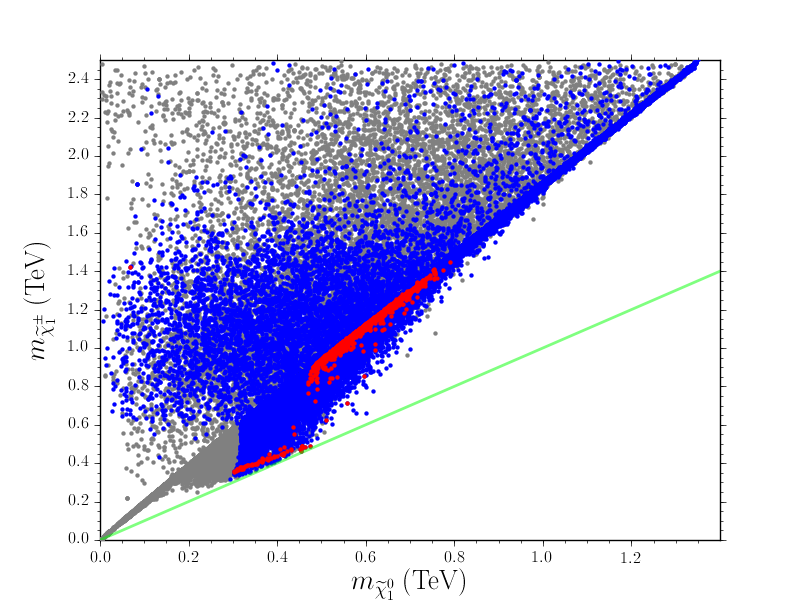}\hspace{0.5cm}
	\includegraphics[scale=0.40]{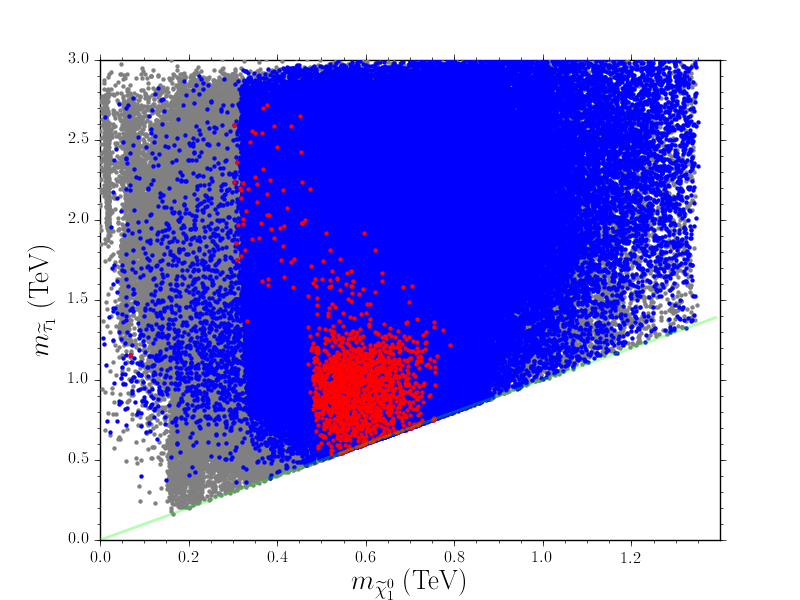}
	\includegraphics[scale=0.40]{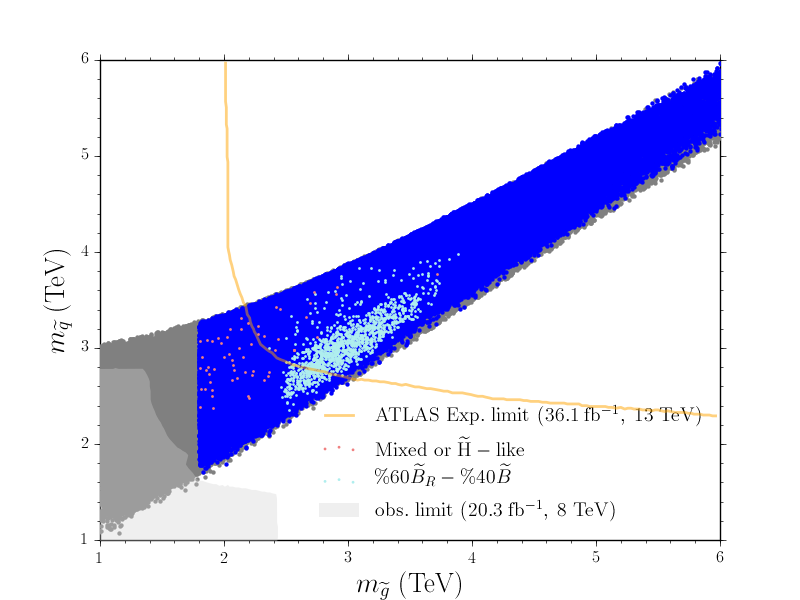}\hspace{0.5cm}
	\includegraphics[scale=0.40]{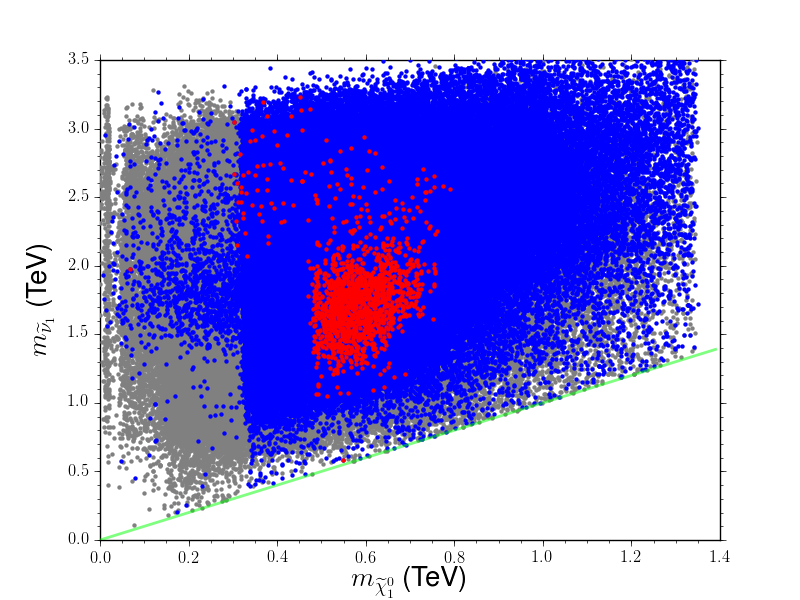}
	\caption{Plots in (top left) $ m_{\widetilde{t}_1} - m_{\widetilde{\chi}_1^0} $, (top right) $ m_{\widetilde{b}_1} - m_{\widetilde{\chi}_1^0} $, (middle left) $ m_{\widetilde{\chi}_1^{\pm}} - m_{\widetilde{\chi}_1^0}$, (middle right) $ m_{\widetilde{\tau}_1} - m_{\widetilde{\chi}_1^0}$, (bottom left) $ m_{\widetilde{q}} - m_{\widetilde{g}} $, and (bottom right) $ m_{\widetilde{\nu}_1} - m_{\widetilde{\chi}_1^0} $ planes.  The color coding is the same as \autoref{fig:freeparams}. In the bottom left panel,  the color coding represents the neutralino composition as indicated in the insert. The solid line in each plane indicates the degenerate mass region. }
	\label{fig:sparticles}
\end{figure}

To continue the investigation of the  neutralino LSP composition,  in \autoref{fig:DMneutralinolsp_2} we plot the correlation  between the neutralino mass and gaugino and higgsino mass ratios with (top left) $ M_4/M_1 $, (top right) $ M_1/\mu $, (bottom left) $ M_2/\mu $,  and (bottom right)  $\mu_R$ - $\mu$, for correct relic density. The color coding is the same as \autoref{fig:freeparams}. According to the $ M_4/M_1 - m_{\widetilde{\chi}_1^0} $ plane, there must be a clear relation between the $B-L$ bino $ \widetilde{B}$ and $ \widetilde{B}_R $ masses so that the ratio of $ \widetilde{B}_R / \widetilde{B} $ should be at around $\sim1.8$, decreasing slightly when the neutralino LSP mass increases. The next two plots compare the bino-higgsino (top right) and wino-higgsino (bottom left) masses, respectively, by looking at their mass ratio. In the top right plot, almost all solutions satisfying LHC collider bounds, and {\it all} solutions satisfying relic density constraints have $ M_1/\mu \lesssim $ 1, that is the bino is lighter than the higgsino mass parameter. The left bottom plane shows that, despite allowing for light higgsinos, the wino is mostly lighter than the higgsino over all the parameter space where relic density bounds are satisfied. The $\mu_R - \mu$ plot (bottom right) shows that solutions prefer positive $ \mu_R $ to the negative ones, and $ \mu_R $ can take values in a large range between   500 GeV--7 TeV while the relic density bound can only be fulfilled with the low $\mu$ values. As can be seen from $\mu_R - \mu$ plane, the  relic density constraint can be satisfied mostly when $\mu \lesssim 0.5 $ TeV and $ 0.7 $ TeV $ \lesssim \mu \lesssim 1.5 $ TeV. 

The neutralino LSP content consistent with all constraints (including relic density)  is  as follows: its mass is constrained as  $ 300 $ GeV $ \lesssim m_{\widetilde{\chi}_1^0} \lesssim 500 $ GeV, and for those parameter points, the neutralino LSP content is  a $\widetilde{B}_R$-ino, $\widetilde{H}$-ino and $\widetilde{B}$-ino mixture,  in this region the  wino masses are heavier than the higgsino masses for solutions consistent with the relic density bound.  Since $ M_1/\mu \lesssim $ 1, the bino mixes more than the higgsinos to form the LSP neutralino. In the region $ 500 $ GeV $ \lesssim m_{\widetilde{\chi}_1^0} \lesssim 800 $ GeV, the LPS neutralino is about $60\% \widetilde{B}_R - 40\% \widetilde{B}$ admixture, consistent also with the top left plot in \autoref{fig:DMneutralinolsp}.
\begin{figure}
	\centering
	\includegraphics[scale=0.40]{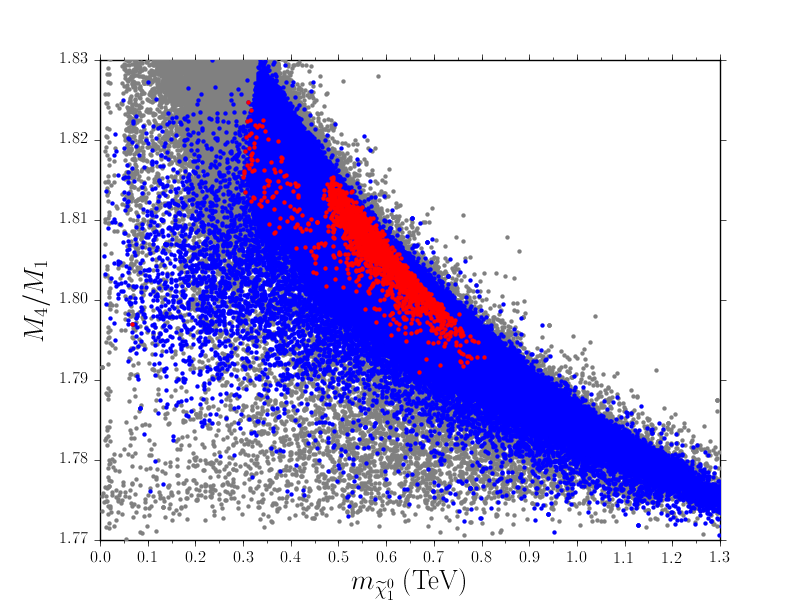}\hspace{0.5cm}
	\includegraphics[scale=0.40]{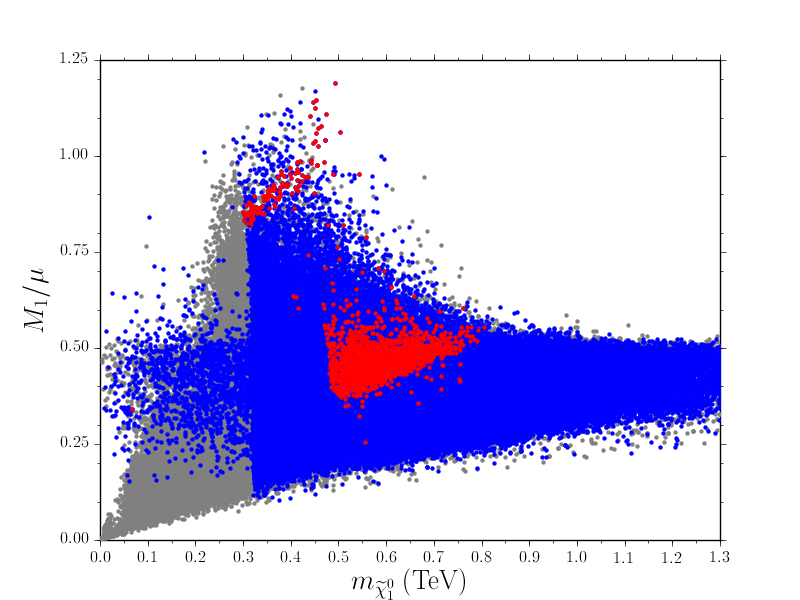}
	\includegraphics[scale=0.40]{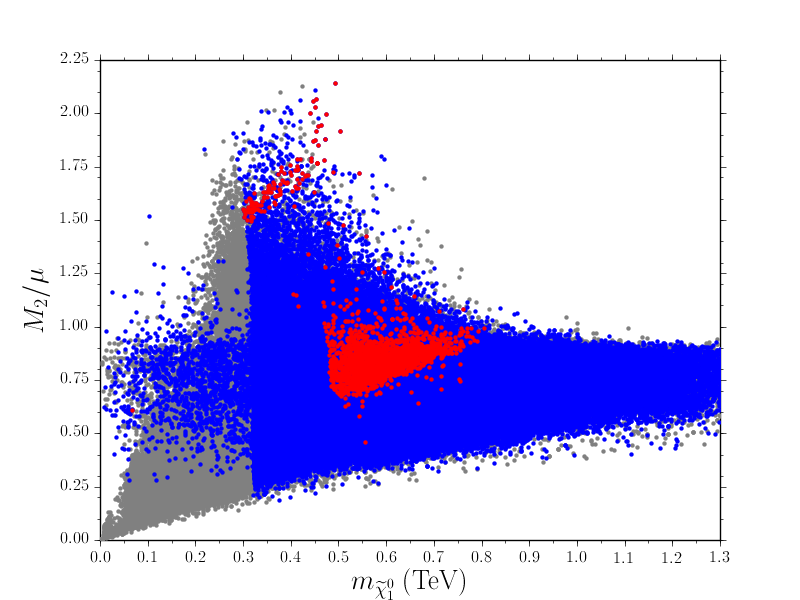}\hspace{0.5cm}
	\includegraphics[scale=0.40]{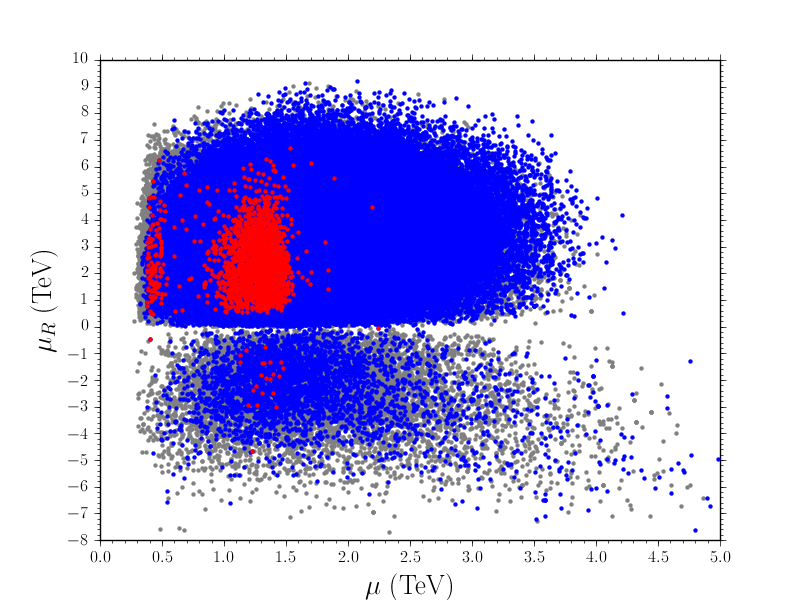}
	\caption{Plots for the neutralino LSP mass and mass ratios: (top left) $ M_4/M_1 $, (top right) $ M_1/\mu $, (bottom left) $ M_2/\mu $,  and (bottom right) $\mu_R$ - $\mu$ correlations.  The color coding is the same as \autoref{fig:freeparams}.}
	\label{fig:DMneutralinolsp_2}
\end{figure}

In \autoref{fig:DMneutralinolsp} we present  results specific to dark matter phenomenology, plotting the relic density and  spin-independent cross section  as a function of the lightest neutralino mass.  In addition, we plot the correlation between the lightest pseudoscalar and the third lightest neutral Higgs boson $h_3$, to highlight the fact that dark matter annihilation proceeds through these two funnels. We show (top left) $\Omega_{DM} h^2 - m_{\widetilde{\chi}_1^0}  $, (top right) $ \sigma_{nucleon}^{SI}- m_{\widetilde{\chi}_1^0}$, (bottom left) $m_{A_1} - m_{\widetilde{\chi}_1^0}  $, and (bottom right) $ m_{h_3} - m_{A_1} $ plots. In the top left and top right plane color coding is indicated in the insert, while for the bottom plots the color coding is the same as \autoref{fig:DMneutralinolsp_2}. The top left plot confirms our previous results on the content of LSP neutralino between 500 -- 800 GeV is composed of  60\% $\widetilde{B}_R$-ino and  40\% $\widetilde{B}$-ino, whereas when 300 GeV $ \lesssim m_{\widetilde{\chi}_1^0} \lesssim $ 500 GeV, its content is shared among $\widetilde{B}_R$-ino, $\widetilde{H}$-ino and $\widetilde{B}$-ino . The top left plot shows the dependence of the relic density, and the right plot shows  the dependence of the spin-independent proton and neutron cross section, with neutralino LSP mass.  The  solid green line represents the current exclusion limit for XENON1T experiment \cite{Aprile:2017iyp}. As can be seen from the graph, most  solutions consistent with the  relic density constraint can be found below the XENON1T exclusion bound, specifically between $10^{-10} $ pb -- $10^{-11}$ pb. Hence they can be detected by the next generation DM detectors such as XENONnT \cite{Aprile:2015uzo}, LZ and DARWIN \cite{Aalbers:2016jon}. Note that we also have a substantial amount of  solutions consistent with the relic density above XENON1T exclusion limit. These solutions correspond to the region where 300 GeV $ \lesssim m_{\widetilde{\chi}_1^0} \lesssim $ 500 GeV and where the LSP content  is the mixture of $\widetilde{B}_R$-ino, $\widetilde{H}$-ino and $\widetilde{B}$-ino. Thus   all solutions surviving consistency with both the current XENON1T exclusion limit and the relic density constraint consist of LSP neutralinos with 500 GeV $ \lesssim m_{\widetilde{\chi}_1^0} \lesssim $ 800 GeV, and  with 60\% $\widetilde{B}_R$ and 40\%  $\widetilde{B}$ admixture. Finally, the $ m_{A_1} -m_{\widetilde{\chi}_1^0}$ and $ m_{h_3} - m_{A_1} $  plots indicate  the funnel channels for the LSP neutralino. The solid green line displays the degenerate mass region for the lightest CP-odd Higgs boson and the LSP neutralino, while the yellow shadowed region indicates  solutions with $ m_{A_1} = 2 m_{\widetilde{\chi}_1^0} $, within 8\% error. As can be seen from the graph, the lightest CP-odd Higgs boson, or the neutral $h_3$ Higgs boson can annihilate into two LSP neutralinos when 450 GeV $ \lesssim m_{\widetilde{\chi}_1^0} \lesssim $ 800 GeV. Solutions consistent with the relic density constraint can be found when $A_1$ is degenerate with $h_3$, with mass between 1 and 3 TeV. In this energy scale  $A_1$ and $h_3$ provide the main funnel channels of this model. Apart from these, we have also verified the relation of the relic density with the IceCube confidence level exclusion and the neutrino flux, and all neutralino LSP solutions surviving relic and cross section bounds are within 1\% confidence level of the experimental result.
\begin{figure}
	\centering
	\includegraphics[scale=0.40]{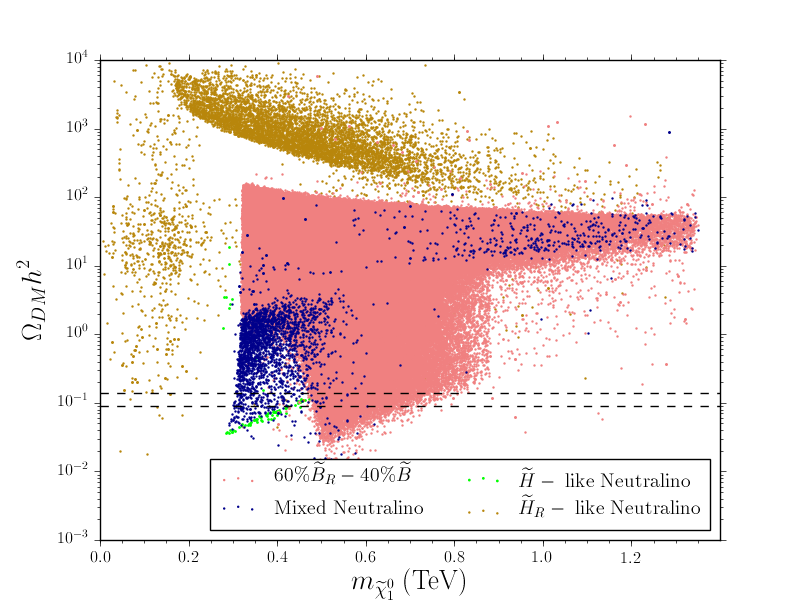}\hspace{0.5cm}
	\includegraphics[scale=0.40]{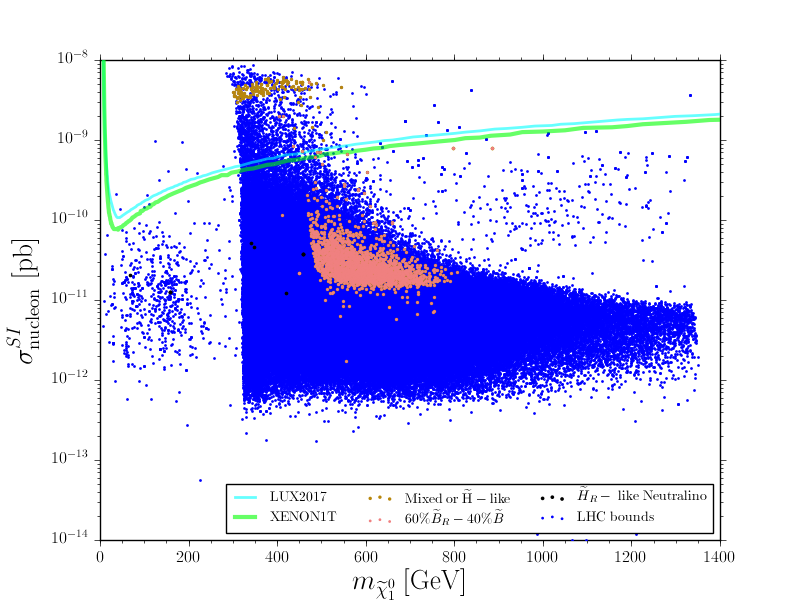}\\
	\includegraphics[scale=0.40]{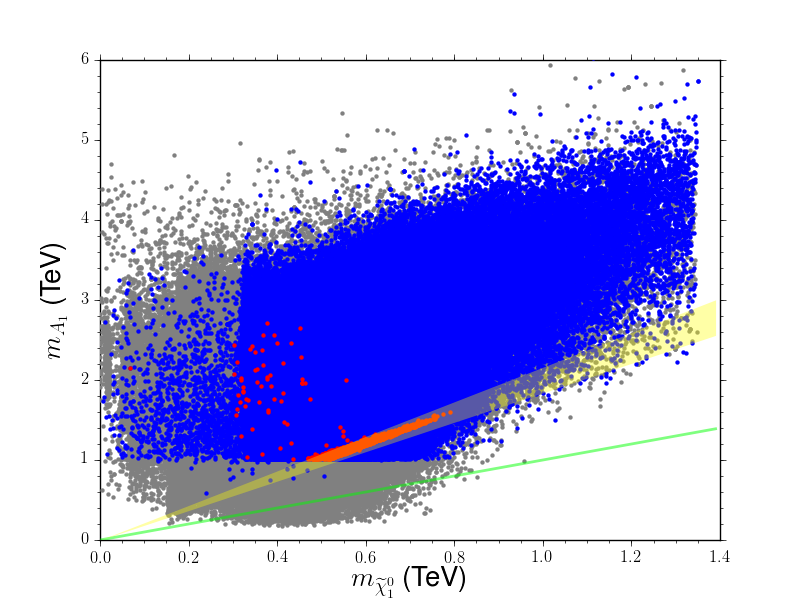}\hspace{0.5cm}
	\includegraphics[scale=0.40]{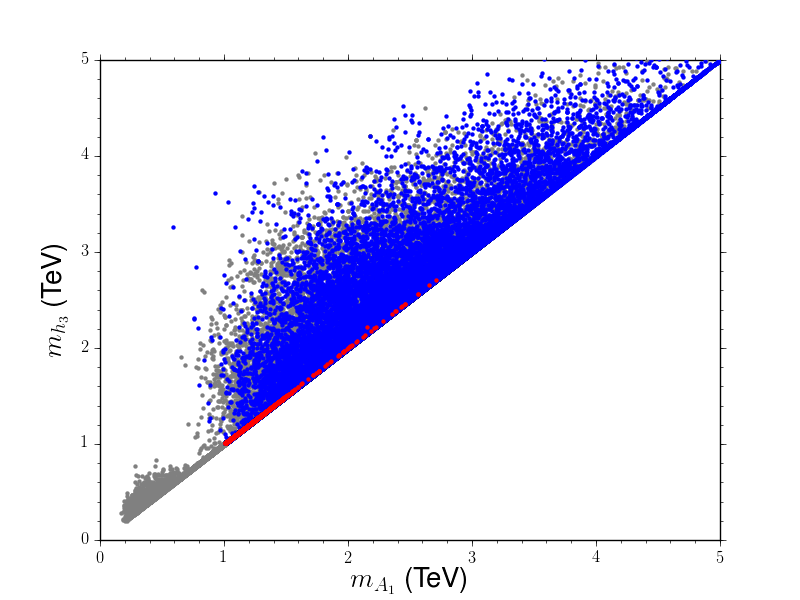}
	\caption{Dependence of: (top left) relic density and (top right) spin independent cross section with nuclei on $ m_{\widetilde{\chi}_1^0} $, (bottom left)  the lightest pseudoscalar Higgs  mass on  $ m_{\widetilde{\chi}_1^0} $ planes, and (bottom right) the degeneracy between the lightest pseudoscalar mass and the third lightest neutral Higgs boson. Both of these provide the funnel channel for the LSP neutralino annihilation.  All points are consistent with LHC and B-physics bounds. The color coding in the $ m_{\widetilde{\chi}_1^0} - m_{A_1} $ plot is the same as in \autoref{fig:freeparams}. The solid line shows the degenerate mass region in these plots.  In addition, the shaded region represents $A_1$ funnel solutions where $ m_{A_1} = 2 m_{\widetilde{\chi}_1^0} $ within 8 \% error.}
	\label{fig:DMneutralinolsp}
\end{figure}

\clearpage

\subsection{Sneutrino Dark Matter}
\label{subsec:sneutrinoDM}

The BLRSSM contains, in addition to the three left sneutrinos, six additional singlet states, three right sneutrinos and three $\widetilde S$, the scalar partners of $S$. The latter two provide candidates for sneutrino dark matter, as they do not suffer from too large an annihilation cross section (thus small relic density) from interacting through $Z$ or $W$ bosons. Sneutrinos thus provide alternative candidates for dark matter in this model, and we analyze their consequences in this subsection.
In the left  and right panels of \autoref{fig:DMsneutrinolsp} we show  the dependence of the relic density $\Omega_{DM} h^2 $ as a function of  the lightest scalar neutrino mass. The color bars to the right of each plot indicate the right-handed sneutrino and the $\widetilde S$ content, respectively. As can be seen from the plot,  even though it is possible to find sneutrino LSP solutions for almost all values of $ m_{\widetilde{\nu}_1} $ between 0--1400 GeV, requiring consistency with the relic density bound constraints LSP sneutrinos to be between 200--400 GeV. Thus the indication would be that sneutrino LSP case allows lighter LSP masses compared to the  neutralino LSP scenario. The right-handed content of the sneutrino LSP solutions changes between 45\%-80\%, while $\widetilde S$ composition varies between 20\%-52\%. Imposing relic density bounds,  the mixed sneutrino LSP is about 50-50 \% between  right-handed and  $\widetilde S$. Thus the scalar partner of $S$, introduced for neutrino seesaw, plays a crucial role in the  sneutrino LSP composition. 

\begin{figure}
	\centering
	\includegraphics[scale=0.42]{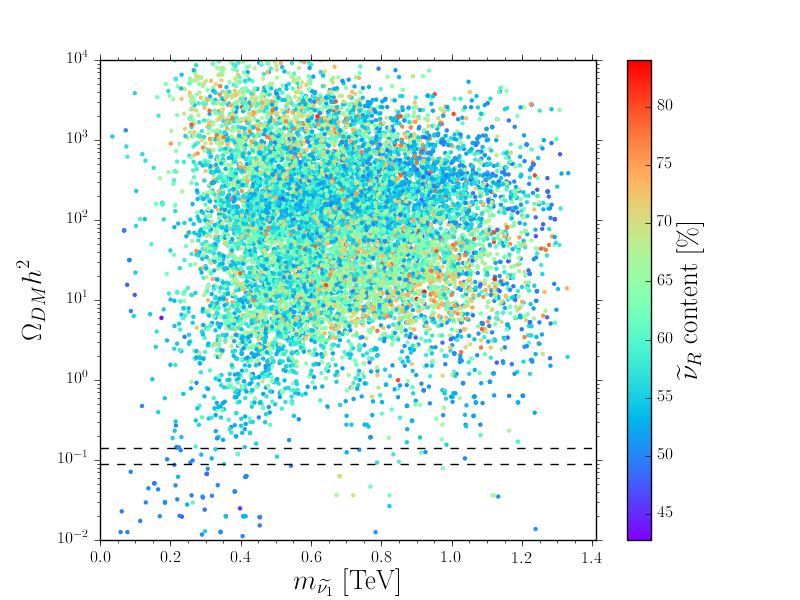}\hspace{0.5cm}
	\includegraphics[scale=0.42]{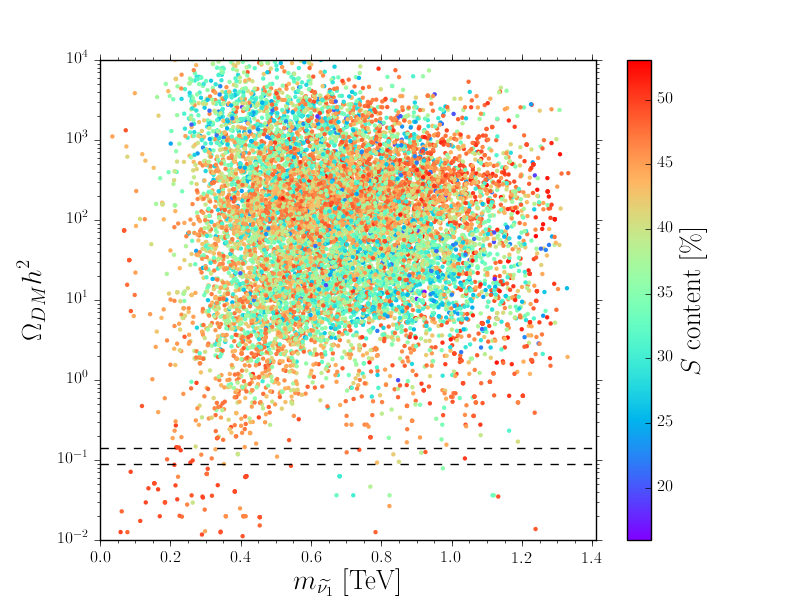}\\
	\caption{ Dependence of  the relic density $  \Omega_{DM} h^2 $ on the lightest sneutrino mass $m_{\widetilde{\nu}_1}$, showing the right sneutrino composition (left panel) and $\widetilde S$ composition (right panel).  All points are consistent with REWSB, LHC bounds, B-physics constraint and sneutrino LSP, while only the points between the two dashed lines satisfy relic density constraints.}
	\label{fig:DMsneutrinolsp}
\end{figure}

In \autoref{fig:DMsneutrinolsp_2} we analyze the dependence of the nucleon spin-independent cross section,  $ \sigma_{p}^{SI} $ for both the proton (left panel) and neutron (right panel). The color coding is the same as \autoref{fig:freeparams} and also indicated in the legend of the plots. The  plots show the relation for the spin independent cross section for proton and neutron respectively. We note that both dark matter constraints (the relic density and $ \sigma_{p}^{SI} $) severely restrict the parameter space where the sneutrino is the LSP in this model.

\begin{figure}
	\centering
	\includegraphics[scale=0.42]{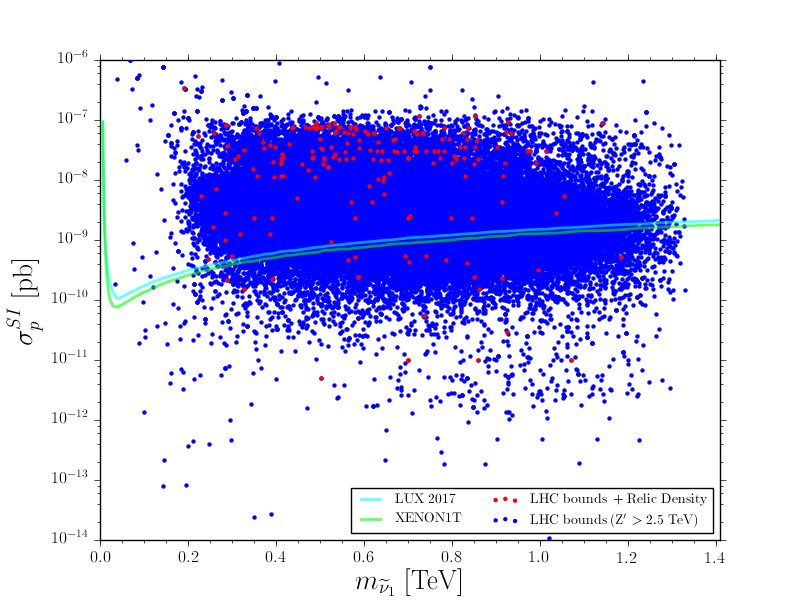}\hspace{0.5cm}
	\includegraphics[scale=0.42]{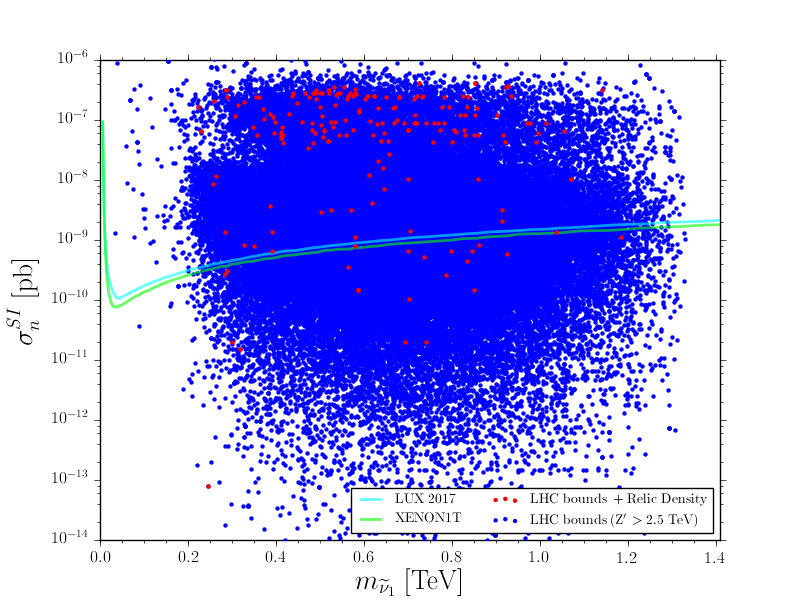}
	\caption{Dependence of the spin independent cross section for the proton $\sigma_{p}^{SI} $ (left) and neutron $ \sigma_{n}^{SI} $ (right) as a function on the sneutrino LSP mass $m_{{\widetilde \nu}_1}$. All points are consistent with REWSB and sneutrino LSP. The color coding in each plane is the same as \autoref{fig:freeparams}. }
	\label{fig:DMsneutrinolsp_2}
\end{figure}

\newpage
\section{Comparison of the two Dark Matter scenarios}
\label{sec:comparisonLSP}

In the previous section, we analyzed  DM phenomenology for both neutralino LSP and sneutrino LSP scenarios in BLRSSM. As discussed in detail, BLRSSM provides quite different mass spectrum for two distinct variants of LSP, and 
these relatively two different mass spectra change the low scale DM phenomenology in important manner. While we found  sneutrino LSP scenario to be highly constrained and  statistically unlikely, there are a few parameter points that survive universal boundary conditions, so in this section, we  compare results for the two different LSP scenarios. In \autoref{fig:electroweakcompare} we plot  
in the $ \mu - \mu_R $ and $ \tan\beta - M_2/\mu $ dependence. Dark blue points satisfy the mass bounds and constraints from the rare $B$-decays for the neutralino LSP solutions. Red points form a subset of dark blue, and represent neutralino LSP solutions which satisfy  the relic density constraint. Light blue solutions are consistent with the mass bounds and the constraints from the rare $B$-decays  for sneutrino LSP solutions, while yellow points form a subset of light blue, and represent sneutrino LSP solutions consistent with the relic density constraint. 

The $ \mu - \mu_R $ plots compare the higgsino sectors of our model. We note that while the neutralino LSP solution can allow values of $\mu_R$ between 7--9 TeV, sneutrino LSP solutions prefer low $\mu_R$ values, mainly between 0--4 TeV for positive $\mu_R$. Even this range becomes narrow, around 1.5 TeV,  for lighter higgsinos. For the sneutrino LSP solutions,  $\mu < 1.5 $ TeV, and $\mu_R$ values favor the region between 4-7 TeV. On the right panel, the $ \tan\beta - M_2/\mu $ plane shows the relative wino and higgsino mass ranges  for the two LSP scenarios.  From the plots, we conclude that for sneutrino LSP, $M_2/\mu \lesssim $ 1 and the wino is always lighter than the higgsino over all the parameter space.  For the neutralino LSP case, the higgsinos can be lighter or heavier than winos. Also, $\tan \beta$ values for sneutrino LSP solutions are found anywhere in the  0--50 range, 
and  solutions consistent with the relic density constraint can be obtained for either   $M_2/\mu \lesssim $ 1 or $M_2/\mu \gtrsim $ 1. Requiring consistency with  the relic density bound solutions with $M_2/\mu \gtrsim $ 1 correspond to neutralino  LSP, and $\tan \beta$ values lie in the 10--50 range. 
 Requiring compatibility  with the relic density bound, further constrains the region $M_2/\mu \lesssim $ 1 to correspond to $\widetilde{B}-\widetilde{B}_R$ dominated neutralino LSP solution,  where $\tan \beta$ should be between 40--60.

\begin{figure}
	\centering
	\includegraphics[scale=0.42]{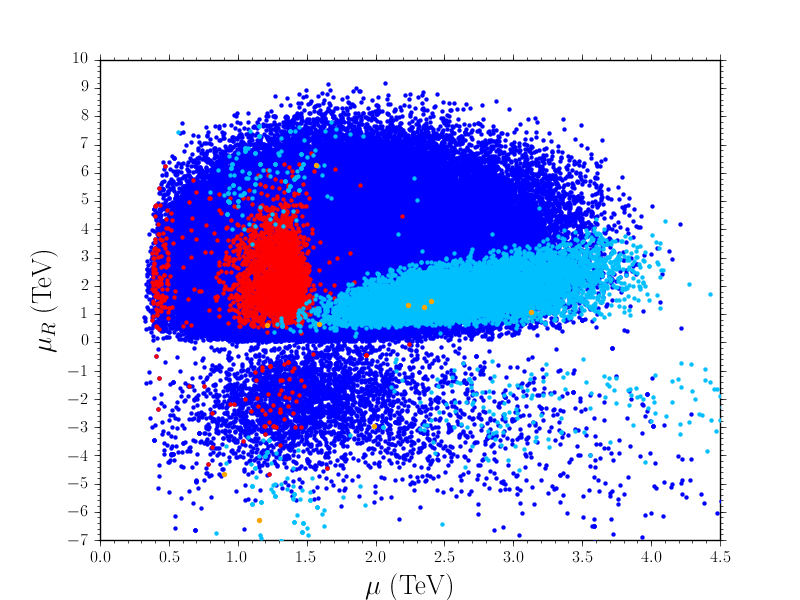}\hspace{0.5cm}
	\includegraphics[scale=0.42]{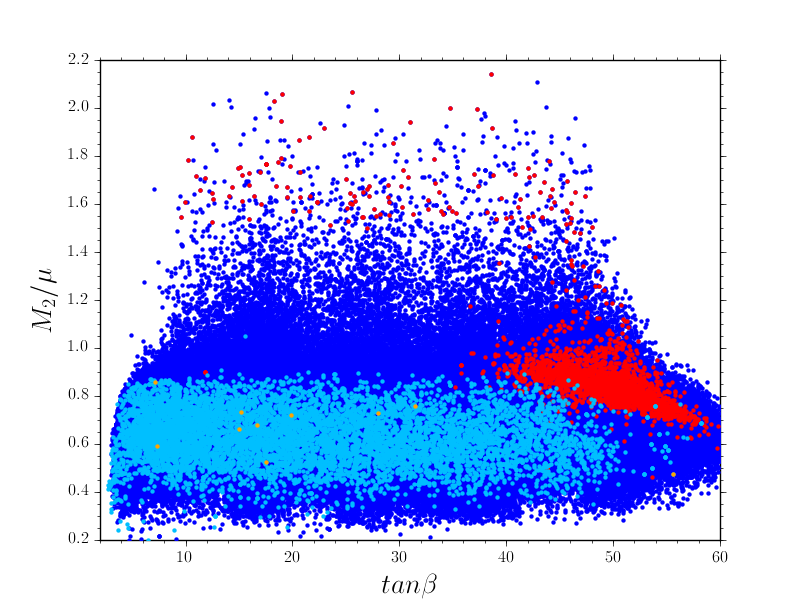}
	\caption{Dependence of higgsino parameters $ \mu_R$ and  $\mu$ (left),  and of $ M_2/\mu $ of $\tan \beta$ (right). All points are consistent with LHC, B-physics bounds, \textsc{HiggsBounds} and \textsc{HiggsSignals}. Dark blue points displays neutralino LSP solutions whereas light blue ones stand for sneutrino LSP solutions. Red points represent the neutralino LSP solution,  while yellow ones stand for sneutrino LSP solutions, consistent in addition, with the relic density bound.}
	\label{fig:electroweakcompare}
\end{figure}

In general the model clearly favors solutions with neutralino LSP to those with sneutrino LSP.

\subsection{The neutral Higgs sector}
\label{subsec:Higgssector}

The choice of LSP affects the heavier states in the Higgs Sector of BLRMSS. For both neutralino and sneutrino LSP solutions, the lightest neutral  Higgs boson can be lighter than 150 GeV. \autoref{fig:higgssec} shows the results for the  values of Higgs masses for both LSP cases with plots for $ m_{h_2}$ relative to $m_{h_1} $ (left) and $ m_{A_1}$ dependence of  $\tan \beta$ (right), where $A_1$ is the lightest pseudoscalar. The color coding is described in the legend of these planes. The left  plot shows that while the two lightest neutral Higgs bosons can be degenerate when the LSP is neutralino, degenerate solutions cannot be obtained for the sneutrino LSP, where the second lightest Higgs boson mass is between 150 -700 GeV. 
This phenomenon can be explained as due the contributions obtained from different elements of CP-even Higgs mass matrix. When $ m_{h_2} > $ 150 GeV, the dominant contribution comes from the $m_{RR}^2$ element of CP-even Higgs mass matrix, corresponding to singlet Higgs fields associated with $U(1)_R \times U(1)_{B-L}$. Thus there $h_2$  is mostly  a singlet Higgs boson.  The off-diagonal term $m_{LR}^2$ which provides essential mixing between the two sectors becomes  important when $ m_{h_2} < $ 150 GeV. For the sneutrino LSP solutions, the Yukawa coupling $Y_s$ is constrained to be small (as the sneutrino LSP mass is generated mostly through this term),  unlike  when the LSP is the neutralino. The $Y_s$ coupling then imposes lighter  $h_2$ masses,  mostly generated by the singlet Higgs field $\singR$. The other Higgs bosons can be quite heavy.  This is seen also in the right-hand side of \autoref{fig:higgssec}, where we plot the dependence of the mass of the lightest pseudoscalar Higgs boson $A_1$ (degenerate with $h_3$), with $\tan \beta$. As before, the region in $\tan \beta \sim$ 40-60 represents the mixed binos neutralino LSP solutions, while for $\tan \beta<40$, regions with larger (smaller) $A_1$ mass correspond to sneutrino (neutralino) LSP. Thus second lightest Higgs boson is a singlet in both scenarios, but, while the sneutrino LSP scenario favors the 150-700 GeV mass range, for the neutralino LSP solutions the second Higgs mass can be
much heavier than 700 GeV.
\begin{figure}
	\centering
	\includegraphics[scale=0.42]{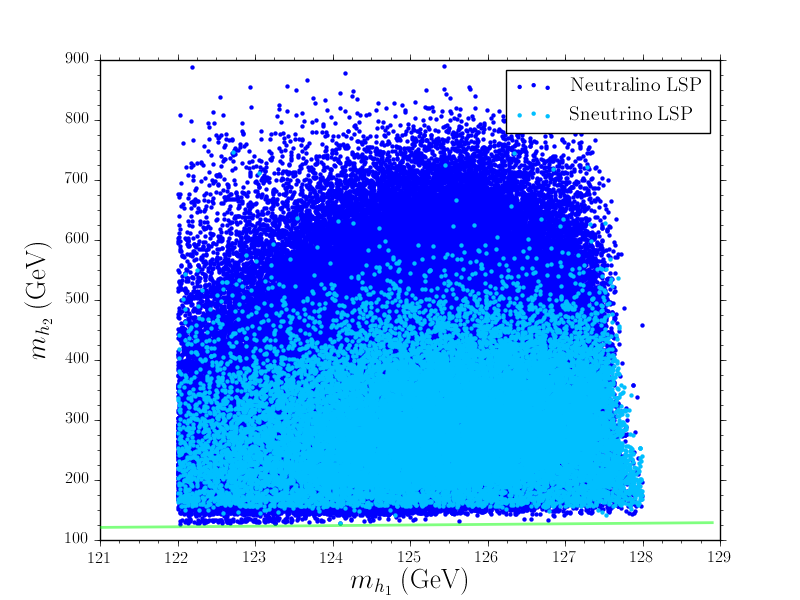}\hspace{0.5cm}
	\includegraphics[scale=0.42]{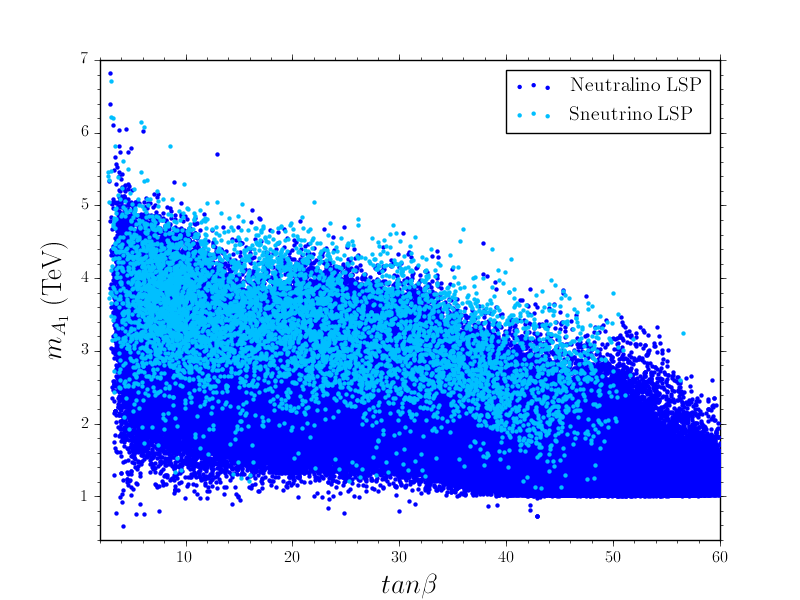}\\
	\caption{Dependence of  $ m_{h_2}$ and $m_{h_1} $ (left)  and dependence of $ m_{A_1}$ on  $\tan \beta $ (right). The color coding is the same as \autoref{fig:electroweakcompare}. In addition, the solid green line shows the degenerate mass region where $ m_{h_1} = m_{h_2} $.}
	\label{fig:higgssec}
\end{figure}

\subsection{The muon anomalous magnetic moment}
\label{subsec:muong2}

The experimental results for the muon anomalous magnetic moment pioneered by the BNL E821 experiment \cite{PhysRevD.73.072003,PhysRevD.80.052008} have been improved with the updated results from FNAL E989 \cite{Grange:2015fou} and J-PARC E34  \cite{Saito:2012zz} experiments. However, the SM prediction for the muon anomalous magnetic moment \cite{Davier:2010nc}, $a_\mu=(g-2)_\mu/2$, indicates a 3.5$\sigma$ deviation from the experimental results, 
\begin{equation}
\Delta{a_\mu} = a_\mu^{\rm exp} - a_\mu^{\rm SM} = (28.7 \pm 8.0) \times 10^{-10} (1\sigma)
\end{equation}
The SM prediction is limited in precision by the evaluation of hadronic vacuum polarization contributions. Calculations exist for the lowest contributions, evaluated using perturbative QCD and experimental cross section data involving $e^+e^-$ annihilation into hadrons. However, the large discrepancy has motivated possible explanations within new physics scenarios. 

In MSSM, if one of the smuons and bino or wino soft masses can be sufficiently light, supersymmetry can ameliorate this discrepancy. However, if the model is required to obey universality conditions at $M_{\rm GUT}$,  obtaining the correct  Higgs boson mass is the greatest challenge to explaining the muon $g-2$ anomaly. We can expect better results from the BLRSSM model since it includes inverse seesaw mechanism and an extra gauge sector. The effect of inverse seesaw mechanism can be read through RGE for the smuons. As can be seen from the last two terms of \autoref{RGslepton},  the Yukawa coupling  $Y_\nu$ helps  decrease the smuon masses at low scales,  as compared to models without inverse seesaw. A similar effect can be read through the RGE of $\mu$ \autoref{RGMu} and sneutrinos \autoref{RGsneutrino}. The presence of another free Yukawa coupling $Y_s$ in addition to $Y_\nu$ contributes to evolving light sneutrino masses to the low scale via RGE as can be seen from the \autoref{RGsneutrino}.

Here we investigate the effects on the  muon $g-2$ anomaly for both sneutrino and neutralino LSP cases. \autoref{fig:muog2} displays the correlations between muon $a_{\mu}$ and the relevant free parameters in  $ m_{0}$, $ M_{1/2} $,  $ \tan \beta $ and $\mu$. The color coding is the same as \autoref{fig:higgssec}. In addition, the shadowed regions show 1, 2  and 3 $\sigma$ deviations between the calculated contribution to muon $g-2$ factor and its experimental value. The top left side plot shows that $\Delta a_\mu$ favors low values for $ m_{0}$ (light scalar masses). Similarly light gaugino masses  (light electroweakinos) are also required to decrease the $\Delta a_\mu$ discrepancy, as seen from  the top right handed plot. The need of light scalars and electroweakinos agrees with large $\tan \beta $ values (bottom left panel). Finally, the $\Delta a_\mu$ depends sensitively on the $\mu$ parameter, as in MSSM, and here the contribution to the muon $g-2$ factor drops sharply for $\mu > $ 1.5 TeV. This is due to one loop contributions effects, where, as the $\mu$ term increases, the contributions where the higgsinos run in the loop  are suppressed, while bino-smuon loop is left as only effective contributing diagram. However, as the 
bino masses cannot be as low as ${\tilde B}_R$ masses, the  contribution from this channel is insufficient. And thus, against expectations, the inverse seesaw mechanism cannot sufficiently enhance muon $\Delta a_\mu$ within universality conditions, and the corrections hardly reach the $2\sigma$ region. 
\begin{figure}
	\centering
	\includegraphics[scale=0.40]{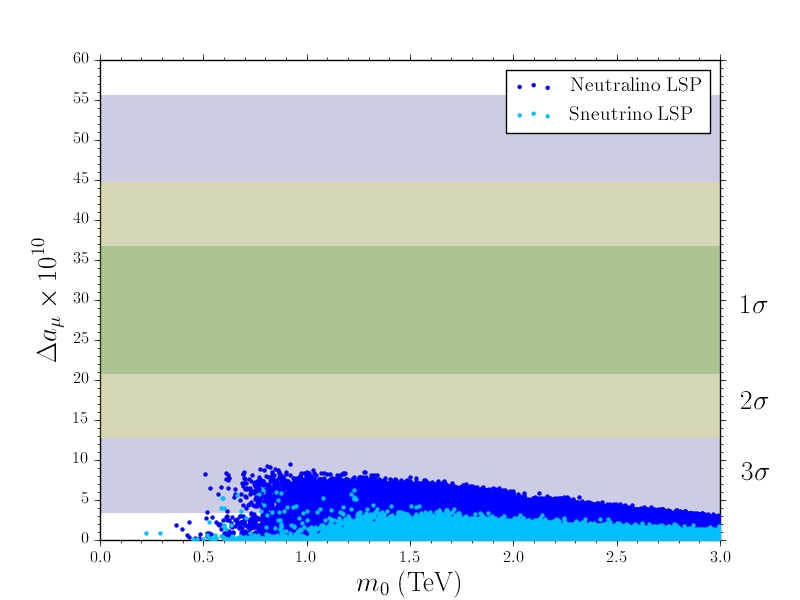}\hspace{0.5cm}
	\includegraphics[scale=0.40]{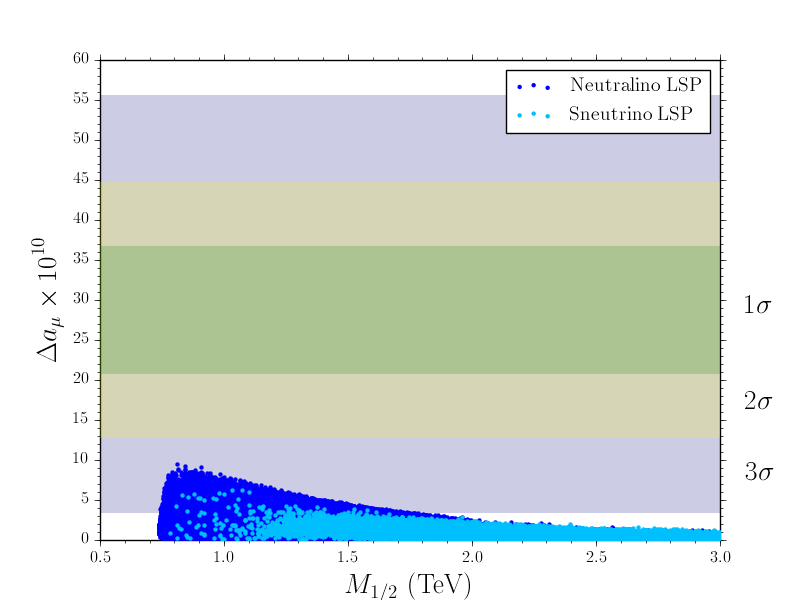}\\
	\includegraphics[scale=0.40]{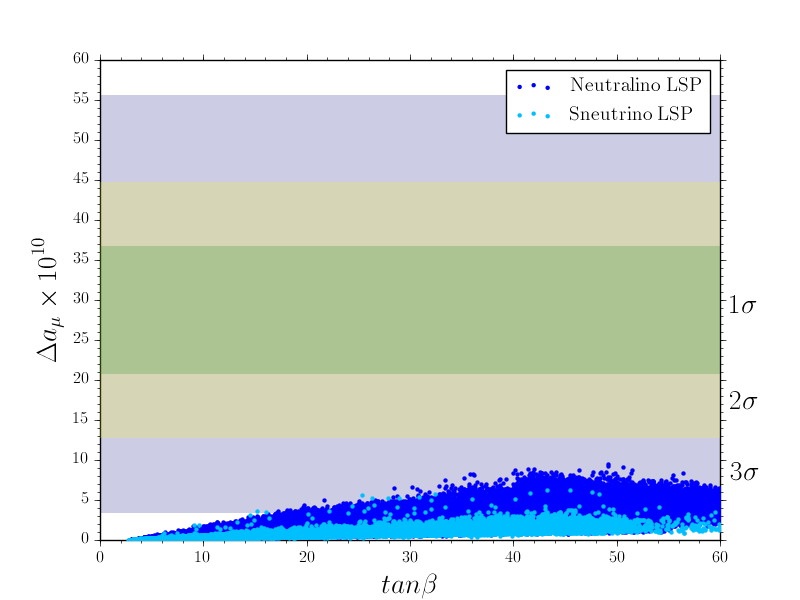}\hspace{0.5cm}
	\includegraphics[scale=0.40]{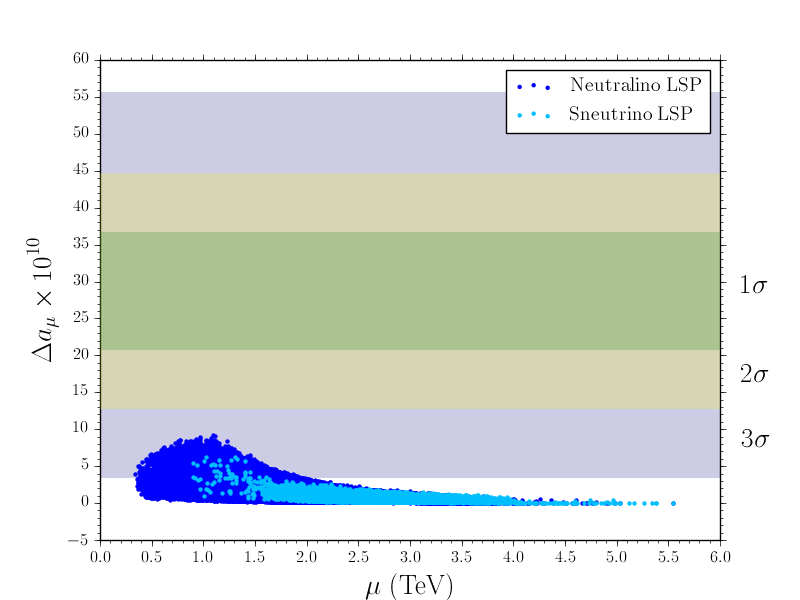}
	\caption{$\Delta a_\mu$ dependence of $ m_{0}$ (top left) , $ M_{1/2}$ (top right), $ \tan{\beta}$ (bottom left) and $ \mu $ (bottom right). 
	 The color coding is the same as \autoref{fig:higgssec}. In addition, the shadowed regions show 1 $\sigma$, 2 $\sigma$ and 3 $\sigma$ differences between the theoretical contribution to muon $g-2$ factor and its experimental value.}
	\label{fig:muog2}
\end{figure}

\subsection{$Z^\prime$ mass constraints}
\label{subsec:zprime}

To highlight the differences between the two scenarios, we kept the model as general as possible  and did not   impose $Z^\prime$ mass bounds so far.  We include here an investigation of implications of the constraints imposed on the   $Z^{\prime}$ mass by a recent new study at ATLAS \cite{ATLAS:2017wce},  requiring an increase in the lower bound for the BLRSSM model to $M_{Z^{\prime}} > 3.9\,(3.6)$ TeV in the $ee (\mu \mu)$ channels. One must be careful when applying these bounds. First, the experiment assumes non-supersymmetric models, and thus a case where $Z^\prime$ does not decay to supersymmetric particles, which will modify its total decay width and thus branching ratios. Second, the parameter choice and unification scale is different from ours: the choice depends on symmetry breaking scales and assumed multiplet composition of the GUT parent. With this note of caution, we explore the parameter space here.

First, we show some of the decay rates of the $Z^\prime$ boson in BLRSSM. \autoref{fig:ZpBR} displays some of the important decay channels of $ Z^{\prime} $  where  $ BR(Z^{\prime} \to ll) = BR(Z^{\prime} \to ee) + BR(Z^{\prime} \to \mu\mu) $, $ BR(Z^{\prime} \to \widetilde{l} \widetilde{l}) $, $  BR(Z^{\prime} \to qq) $ and $  BR(Z^{\prime} \to \widetilde{\chi} \widetilde{\chi}) $, all plots as a function of $m_{Z^\prime}$. Throughout, all points are consistent with LHC, B-physics bounds, \textsc{HiggsBounds} and \textsc{HiggsSignals}. Dark blue points show neutralino LSP solutions whereas light blue ones stand for sneutrino LSP solutions. 

The top left panel in \autoref{fig:ZpBR} exhibits the branching ratio of $ Z^{\prime} $ into lepton pairs while the top right panel shows the branching for the supersymmetric partners in the same channel. As can be seen from top left plane, the branching ratio of $ Z^{\prime} $ into leptons changes between $25\%-37\%$ while its decays into their supersymmetric partners, sleptons, are low, in the range of 0\% and 8\%. It is interesting to note that these models, unlike $E_6$-derived models containing an extra $U(1)^\prime$ gauge group,  are not likely to be leptophobic as the branching ratio into leptons is significant throughout the parameter space investigated. The bottom panels of  \autoref{fig:ZpBR} show the branching ratio into quarks (left) and into neutralinos and/or charginos (right). As usual, the largest branching ratio obtained is hadronic (40\%-62\%), which, though significant, is not as large as for $U(1)^\prime$ models \cite{Araz:2017qcs}, which will likely adversely affect the $Z^\prime$ production cross section. The decay into two charginos or neutralinos occurs above their mass threshold and is very small throughout the whole parameter space (0\%-13\%). So it appears that the decay of the $Z^\prime$ boson is fairly consistent with a non-supersymmetric scenario. Based on this, we shall investigate the effects of setting the mass lower bound to be $ m_{Z^{\prime}} > 3.5 $ TeV throughout our analyses.
\begin{figure}
	\centering
	\includegraphics[scale=0.42]{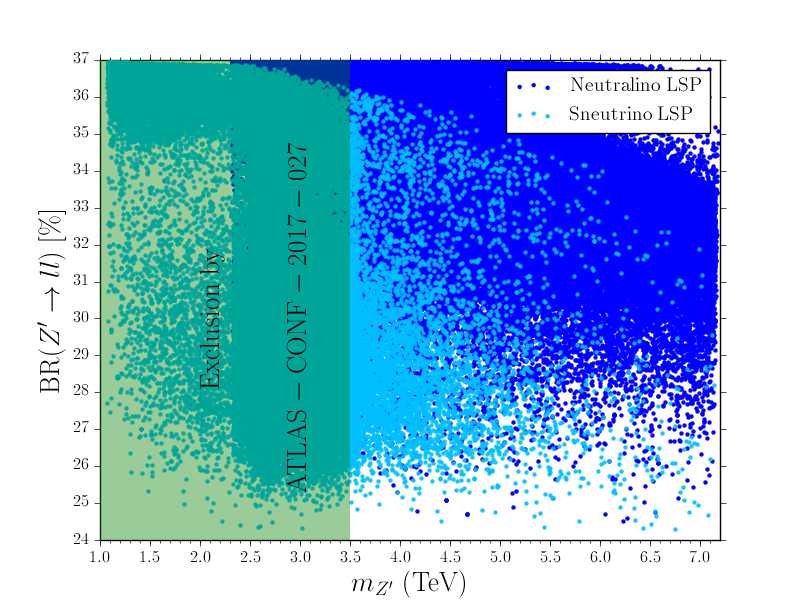}\hspace{0.5cm}
	\includegraphics[scale=0.42]{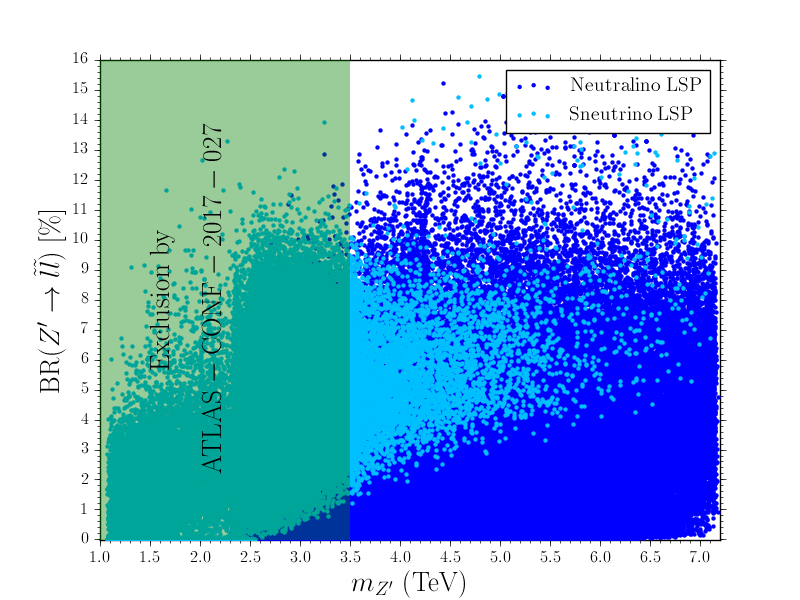}\\
	\includegraphics[scale=0.42]{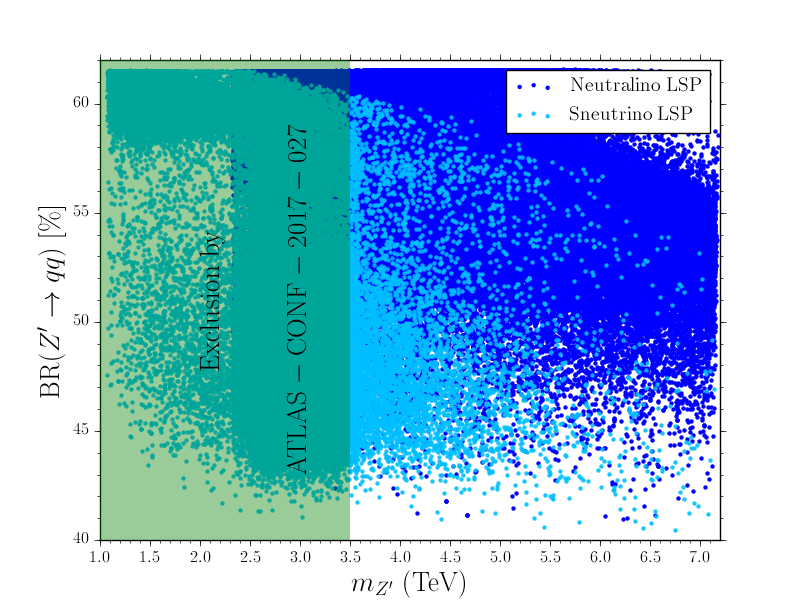}\hspace{0.5cm}
	\includegraphics[scale=0.42]{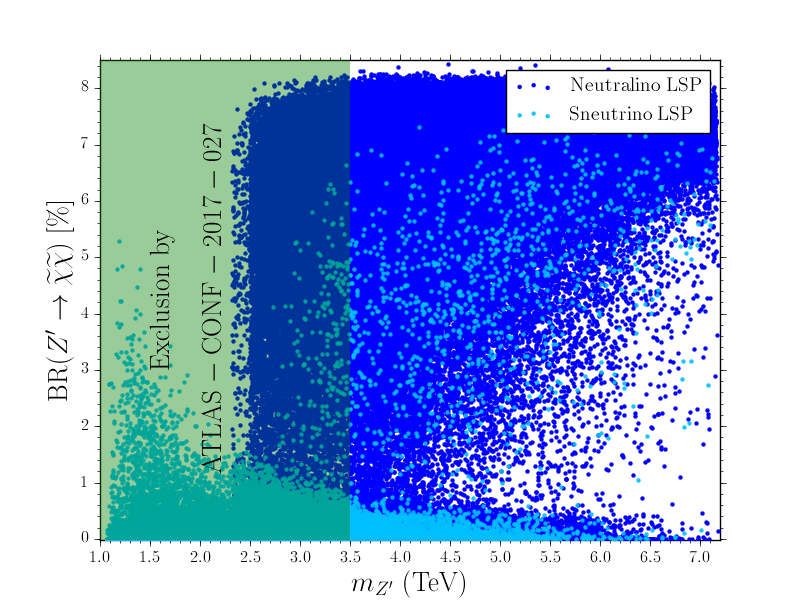}
	\caption{ Branching ratios of $Z^\prime$ in BLRSSM.  (Top left): $BR(Z^{\prime} \to ll (ee+\mu \mu)) $; (top right) $BR(Z^{\prime} \to \widetilde{l} \widetilde{l}) $, (bottom left) $BR(Z^{\prime} \to q\bar{q}) $ and  (bottom right) $BR(Z^{\prime} \to \widetilde{\chi} \widetilde{\chi}) $. Neutralino LSP points are represented in dark blue, sneutrino LSP points in light blue. The solutions excluded by ATLAS-CONF-2017-027 are  in the shaded green region.}
	\label{fig:ZpBR}
\end{figure}

With these constraints, we revisit the plots for the spin independent cross section for proton and neutron respectively. 
While in the \autoref{fig:DMsneutrinolsp_2}, we considered  $m_{Z^\prime} \ge 2.5$ TeV, and the
spin-independent proton (or neutron) cross sections for sneutrino LSP solutions were satisfied with XENON1T experimental exclusion limit, imposing the the new $Z^\prime $ mass limit excludes most of the parameter space for sneutrino LSP solutions, as shown in \autoref{fig:DMsneutrino_zprime}. Specifically, of about $10^6$ scanned parameter points only 18 solutions compatible with the relic density bound are found, and only 10 of them can survive XENON1T experimental exclusion limit. Imposing  $Z^\prime$ mass constraints, the sneutrino LSP scenario thus emerges as extremely constrained and, realistically, ruled out. 

\begin{figure}
	\centering
	\includegraphics[scale=0.42]{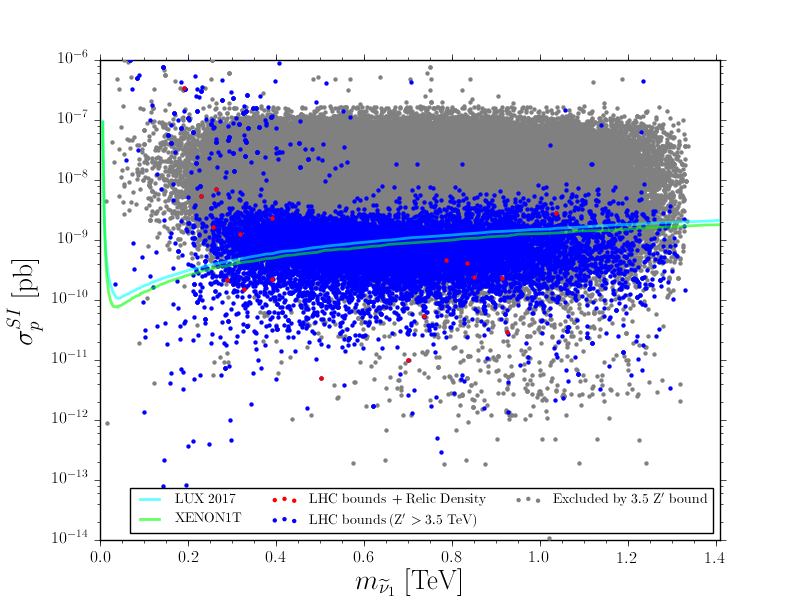}\hspace{0.5cm}
	\includegraphics[scale=0.42]{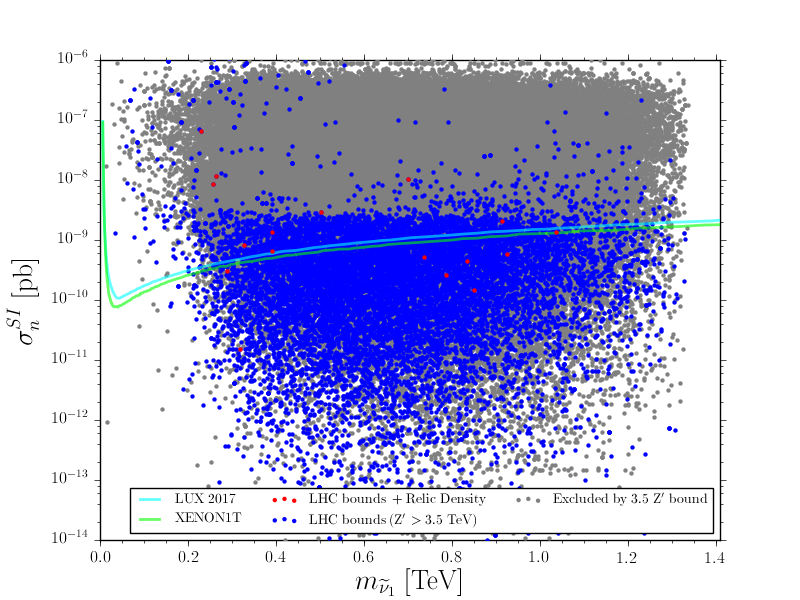}
	\caption{Dependence of the spin independent cross section for the proton $\sigma_{p}^{SI} $ (left) and neutron $ \sigma_{n}^{SI} $ (right) as a function on the sneutrino LSP mass $m_{{\widetilde \nu}_1}$, for , $m_{Z^{\prime}} \geq$ 3.5 TeV. All points are consistent with REWSB and sneutrino LSP. The color coding in each plane is the same as \autoref{fig:freeparams}. }
	\label{fig:DMsneutrino_zprime}
\end{figure}

\section{Collider Signals}
\label{sec:collider}

Lastly, we would like to  analyze the production and decays for this scenario at the LHC.   We choose benchmarks from the parameter scan results which satisfy all experimental bounds, including the relic density constraint and XENON1T exclusion limits, and favor light neutralino LSP solutions as  the only ones surviving all constraints. We proceed by exporting the BLRSSM to the UFO format \cite{Degrande:2011ua} and use $\rm {MG5\_aMC@NLO}$ framework version 2.5.5 \cite{Alwall:2014hca} to simulate hard-scattering LHC collisions and evaluate the cross sections for various signals. For the calculation of cross sections, we select four  benchmarks with different features, which could showcase different features of the model for detection at the LHC. 

The first benchmark,  benchmark 1  has $H_{\widetilde{R}}$-like neutralino LSP.  (Even though parameter scans allow Higgsino-like and higgsino-binos mixed LSP neutralino solutions between 300-500 GeV, no benchmark in this range can be found as these states are completely excluded by the XENON1T exclusion limit.)  We thus select benchmarks with mixed $B_{\widetilde{R}}-\widetilde{B}$ content. For benchmarks 2-3,  BR$(\widetilde{\chi}_2^0 \to \widetilde{\chi}_1^0 h_1) $ and BR$(\widetilde{\chi}_1^\pm \to \widetilde{\chi}_1^0 W^\pm) $ are almost  unity. Sparticle masses are similar in both cases, with the exception of the lightest charging, which is heavier  for benchmark 3. Also, for benchmark 3,  BR$(\widetilde{\tau}_1 \to \tau_1 \widetilde{\chi}_1^0) \sim1$ while this is much smaller for benchmark 2.  Benchmark 4 is selected for light stau masses, leading to increased  stau-stau production cross sections.  Note that benchmarks satisfy all the constraints, including  the Icecube22 exclusion.  Our results are shown in Table \autoref{tab:cross-section}.

Even though LSP neutralino mass is quite light (67 GeV) for benchmark 1, we find that both chargino-chargino and neutralino-chargino production cross sections are quite low, due to the fact that the neutrino is mostly higgsino.  For the other benchmarks, with neutralino contents of mixed binos, the second lightest neutralino and chargino  masses are degenerate. We estimated the cross sections  for chargino/neutralinos and stau production as being the most promising. The highest cross-section values for chargino-chargino production and chargino-neutralino production are obtained for benchmark 2 whose neutralino and chargino masses are 470 GeV and 767 GeV, respectively. As can be seen from the \autoref{tab:cross-section}, chargino-chargino production and neutralino-chargino production cross sections are 4.623 fb and 2.249 fb, respectively. The cross-section values decrease in benchmark 3 (with respect to benchmark 2) when neutralino and chargino masses are 506 GeV and 954 GeV (versus 470 and 767 GeV), respectively. Finally, the last benchmark is selected to enhance $\sigma(pp \to \widetilde{\tau}_1 \widetilde{\tau}_1)$ where each stau can decay into a tau and a LSP neutralino $(\widetilde{\tau}_1 \to \tau_1 \widetilde{\chi}_1^0)$. Note that here the stau is NLSP,  and that the branching ratio of stau into a tau and a LSP neutralino is 1. The relevant cross-section is 0.5713 fb, a factor of 10 larger than for benchmarks 2 and 3, but  still too small to be observed at the LHC.  For all benchmarks,  $Z^{\prime}$ masses are above 4 TeV,  consistent with the latest ATLAS result.  Note that gluino masses are about 2.5 TeV for benchmarks 2, 3 and 4, making gluino results testable  at the HL-LHC or by the next generation colliders \cite{Baer:2017yqq,Baer:2016wkz}. 

Including all the constraints, we conclude that production of supersymmetric particles in BLRSSM fall below detector sensitivity. Especially because the final signals will have even lower production cross sections, as they will be suppressed by branching ratios of chargino/neutralinos to missing energy + leptons. A way to improve our results is to relax some or most universality constraints, and looking for effective cuts which would enhance the signal over the background. We shall return to this in a  future work.

\begin{table}[]
	\centering
	\label{tab:cross-section}
	\begin{tabular}{|c|c|c|c|c|}
		\hline
		                               & Benchmark 1                & Benchmark 2                                 & Benchmark 3               & Benchmark 4  \\ \hline
	$m_0$ [GeV]	                       & 1960                       & 1831.4                                      & 2073                      & 2099.4       \\ \hline
	$M_{1/2}$ [GeV]                    & 1723.7                     & 1092.4                                      & 1166.8                    & 1285.8       \\ \hline 
	$tan\beta$                         & 50.9                       & 45.3                                        & 58.6                      & 53.6         \\ \hline
	$A_0$ [GeV]                        & 2816.4                     & 826.5                                       & 1652.8                    & 4972.9       \\ \hline
	$<v_R>$ [GeV]                      & 11611                      & 11711                                       & 12969                     & 11283        \\ \hline
	$Y_\nu \hspace{0.3cm} (M_{\rm SUSY})$& 0.50716                  & 0.0021115                                   & 0.24648                   & 0.47416      \\ \hline
	$Y_s \hspace{0.3cm} (M_{\rm SUSY})$& 0.50672                    & 0.62476                                     & 0.49036                   & 0.59198      \\ \hline
	$\mu$ [GeV]                        & 2245.1                     & 787.67                                      & 1305.67                   & 2191.82      \\ \hline
	$\mu_R$ [GeV]                      & -64.2                      & 1144.9                                      & 3817.63                   & 4473.89      \\ \hline
	$m_{\widetilde{\chi}_1^0}$	[GeV]  & \textbf{67} ($\widetilde{H}_R$-like)& \textbf{470} (mixed $\widetilde{B}_R-\widetilde{B}$) & \textbf{506} (mixed $\widetilde{B}_R-\widetilde{B}$)          &  \textbf{523} (mixed $\widetilde{B}_R-\widetilde{B}$)  \\ \hline
	$m_{\widetilde{\chi}_2^0}$	[GeV]  & \textbf{757}               & \textbf{768}                                & \textbf{954}              & \textbf{983} \\ \hline
	$m_{\widetilde{\chi}_1^\pm}$ [GeV] & \textbf{1421}              & \textbf{767}                                & \textbf{954}              & \textbf{983} \\ \hline
	$m_{h_2}$ [GeV]                    & 358                        & 380                                         & 224                       & 239      \\ \hline
	$m_{h_3}$ [GeV]                    & 2221                       & 1018                                        & 1013                      & 1049     \\ \hline
	$m_{A_1}$ [GeV]                    & 2149                       & 1020                                        & 1017                      & 1051     \\ \hline
	$m_{\widetilde{t}_1}$ [GeV]        & 2893                       & 1977                                        & 2209                      & 2057     \\ \hline
	$m_{\widetilde{b}_1}$ [GeV]	       & 3257                       & 2279                                        & 2458                      & 2290     \\ \hline
	$m_{\widetilde{\tau}_1}$ [GeV]	   & 1154                       & 1332                                        & 1064                      & 559      \\ \hline
	$m_{\widetilde{\nu}_1}$	[GeV]      & 1972                       & 2036                                        & 1858                      & 1332     \\ \hline
	$m_{Z^{\prime}} $	[GeV]          & 4157                       & 4182                                        & 4632                      & 4090     \\ \hline
	$m_{\widetilde{g}}$	[GeV]          & 3720                       & 2473                                        & 2634                      & 2675     \\ \hline
	$\Omega h^2 $                      & 0.112621                   & 0.103201                                    & 0.096158                  & 0.090515 \\ \hline
	$\sigma^{SI}_{nucleon}$ [fb]       & 2.0771$\times 10^{-11}$    & 1.80137 $\times 10^{-10}$                   & 2.43005 $\times 10^{-11}$ & 2.40945 $\times 10^{-11}$  \\ \hline
	Icecube22  Exclusion CL $[\%]$     & 0.014308                   & 0.607672                                    & 0.029368                  & 0.029803  \\ \hline
	$\sigma (pp \to \widetilde{\chi}_1^{\pm} \widetilde{\chi}_2^0) $ [fb]     & 0.001017 & 2.249    & 1.543    &  1.104    \\ \hline
	$\sigma (pp \to \widetilde{\chi}_1^{+} \widetilde{\chi}_1^{-}) $ [fb]     & 0.3289   & 4.623    & 2.598    &  1.941    \\ \hline
	$\sigma (pp \to \widetilde{\tau}_1 \widetilde{\tau}_1) $ [fb]             & 0.0799   & 0.05059  & 0.06468  & 0.5713    \\ \hline
	BR$(\widetilde{\chi}_2^0 \to \widetilde{\chi}_1^0 h_1) $                  & -        & 0.936403 & 0.885733 & 0.133714  \\ \hline
	BR$(\widetilde{\chi}_1^\pm \to \widetilde{\chi}_1^0 W^\pm) $              & -        & 0.998671 & 0.992318 & 0.151927  \\ \hline
	BR$(\widetilde{\tau}_1 \to \tau_1 \widetilde{\chi}_1^0) $                 & -        & 0.512641 & 0.991196 & 1.00      \\ \hline  
	\end{tabular}
	\vskip0.1in
	\caption{Benchmarks for BLRSSM with relevant cross-sections and branching ratios. In bold, the lightest chargino and the two lightest neutralino states.}
    \label{crosssections}	
\end{table}

\section{Summary and Conclusion}
\label{sec:conclusion}

We analyzed the predictions of the mass spectrum in the BLRSSM framework with universal boundary condition,  highlighting the solutions consistent with the DM restrictions (relic density and spin independent cross sections with nucleons) for both neutralino and sneutrino LSP scenarios. We found that the stop and sbottom masses are between 2-3 TeV, and the chargino can be degenerate with the LSP neutralino between 300-500 GeV. In addition, the relic density constraint can be satisfied for masses in the range 300 $ \lesssim m_{\chi_1^0} \lesssim $ 800 GeV. $\widetilde{H}_R $ dominated LSP neutralino solution can be obtained below 300 GeV,  however these solutions  are ruled out by the XENON1T spin independent cross section exclusion curve. When all DM constraints are taken into account, the model favours LSP neutralinos with masses between 500 $ \lesssim m_{\chi_1^0} \lesssim $ 800 GeV, bino-dominated,  and with composition  60\% $\widetilde{B}_R$ - 40\% $ \tilde B$.  We also that,  when the LSP is neutralino, $A_1$ and $h_3$ are funnel channels for pair-producing them.

In addition, the model allows in principle a sneutrino LSP where its content can be either right-handed dominated or mixed, ${\widetilde \nu}_R$ and $\widetilde S$, with masses between 250-1300 GeV. In this sense, sneutrino LSP solutions can be lighter than the neutralino LSP ones. Purely right-handed dominated sneutrino LSP solutions have difficulty to satisfy the relic density constraint and only mixed ones survive. In addition most of the sneutrino LSP solutions are consistent with  the XENON 1T spin independent cross section exclusion curve.  However, strict imposition of the $Z^\prime$ mass bounds basically rule out  the sneutrino solutions, while not having any effect on the neutralino LSP parameter space. This is one of the most important predictions of the model.

The parameter spaces corresponding to neutralino and sneutrino are quite different. If allowed, sneutrino LSP solutions favor low singlet higgsino mass parameter, $\mu_R$, and the second lightest neutral Higgs boson a singlet, while  neutralino LSP favor larger $\mu_R$ parameters. Sneutrino LSP solutions are spread out over the whole range of $\tan \beta$, while neutralino solutions are restricted in the $40 \lesssim \tan \beta \lesssim 60$. Neutralino LSP solutions allow for degenerate masses of the two lightest neutral Higgs bosons, while the sneutrino LSP, although favoring a light $m_{h_2}$, does not. The anomaly in the anomalous magnetic moment of the muon favors neutralino LSP contributions, where for a large range of scalar masses, and a more restricted one for gauginos and higgsinos, the corrections are within $2 \sigma$ of the experimental result, while sneutrino LSP solutions can at best produce results within $3 \sigma$ of the desired values.

We analyzed collider signatures of this proposed scenario, including all constraints, and they are not promising. The largest cross sections are obtained for chargino/neutralino production, and they are at most of ${\cal O}(4)$ fb, without including cascade decays into leptons which would reduce them further. In the future, collider signals could be enhanced by relaxing some of the severe constraints on the model, such as the universality conditions, and finding suitable cuts to enhance signal versus background. This may extend the parameter space, allowing neutrino LSP back into the consideration. Work in these directions is underway.
\begin{acknowledgments}
  Part of numerical calculations reported in this paper was performed using the National Academic Network and Information Center (ULAKBIM) of TUBITAK, High Performance and Grid Computing Center (TRUBA resources), and using  
  High Performance Computing (HPC), managed by Calcul Qu{\'e}bec and Compute Canada. MF acknowledges NSERC for partial financial support under grant number SAP105354.
\end{acknowledgments}

\appendix
\section{Renormalization Group Equations}

We gather below some of the relevant equation referred to in the paper.

\begin{multline}
	\beta_{\mu}^{(1)} =  
	\mu \Big(-3 g_{L}^{2}  + 3 \mbox{Tr}\Big({Y_d  Y_{d}^{\dagger}}\Big)  + 3 \mbox{Tr}\Big({Y_u  Y_{u}^{\dagger}}\Big)  - g_{R}^{2}  - g_{RB}^{2}  + \mbox{Tr}\Big({Y_e  Y_{e}^{\dagger}}\Big) + \mbox{Tr}\Big({Y_v  Y_{v}^{\dagger}}\Big)\Big)\,.
\label{RGMu}
\end{multline}

\begin{multline}
\beta_{\mu_R}^{(1)}=  
-\frac{1}{2} \mu_R \Big(2 g_{R}^{2}  + 2 g_{RB}^{2}  -2 \sqrt{6} g_{BL} g_{RB}  -2 \sqrt{6} g_{BR} g_R  + 3 g_{BL}^{2}  + 3 g_{BR}^{2}  -2 \mbox{Tr}\Big({Y_s  Y_{s}^{\dagger}}\Big)\Big)\,.
\label{RGmuR}
\end{multline}

\begin{multline}
\beta_{B_{\mu}}^{(1)}  =  
+B_{\mu} \Big(-3 g_{L}^{2}  + 3 \mbox{Tr}\Big({Y_d  Y_{d}^{\dagger}}\Big)  + 3 
\mbox{Tr}\Big({Y_u  Y_{u}^{\dagger}}\Big) - g_{R}^{2}  - g_{RB}^{2}  + \mbox{Tr}\Big({Y_e  Y_{e}^{\dagger}}\Big) + \mbox{Tr}\Big({Y_v  Y_{v}^{\dagger}}\Big)\Big) \\
+2 \mu \Big(2 g_R g_{RB} M_{B R}  + 3 g_{L}^{2} M_2  + 3 \mbox{Tr}\Big({Y_{d}^{\dagger}  T_d}\Big)  + 3 \mbox{Tr}\Big({Y_{u}^{\dagger}  T_u}\Big)  + g_{R}^{2} M_4  + g_{RB}^{2} M_1  + \mbox{Tr}\Big({Y_{e}^{\dagger}  T_e}\Big) + \mbox{Tr}\Big({Y_{v}^{\dagger}  T_\nu}\Big)\Big)\,.
\label{RGBmu}
\end{multline}

\begin{multline}
\beta_{B_{\mu_R}}^{(1)} =  
+B_{\mu_R} \Big(-\frac{3}{2} g_{BL}^{2}  -\frac{3}{2} g_{BR}^{2}  - g_{R}^{2}  - g_{RB}^{2}  + \sqrt{6} g_{BL} g_{RB}  + \sqrt{6} g_{BR} g_R  + \mbox{Tr}\Big({Y_s  Y_{s}^{\dagger}}\Big)\Big) \\ 
+\mu_R \Big(3 g_{BL}^{2} M_1 +2 g_{RB}^{2} M_1 -2 \sqrt{6} g_{BR} g_{RB} M_{B R} +4 g_R g_{RB} M_{B R} -2 g_{BL} \Big(-3 g_{BR} M_{B R}  + \sqrt{6} g_{RB} M_1  + \sqrt{6} g_R M_{B R} \Big) \\ 
+3 g_{BR}^{2} M_4 -2 \sqrt{6} g_{BR} g_R M_4 +2 g_{R}^{2} M_4 +2 \mbox{Tr}\Big({Y_{s}^{\dagger}  T_{s}}\Big) \Big)\,.
\end{multline}

\begin{multline}
\beta_{m_{H_u}^2}^{(1)} =  
-2 g_{R}^{2} |M_4|^2 -6 g_{L}^{2} |M_2|^2 -2 g_{RB} \Big(g_{RB} M_1  + g_R M_{B R} \Big)M_1^* -2 \Big(g_{R}^{2} M_{B R}  + g_{RB}^{2} M_{B R}  + g_R g_{RB} \Big(M_1 + M_4\Big)\Big)M_{B R}^* \\ 
-2 g_R g_{RB} M_{B R} M_4^* +g_{RB} \sigma_{1,1} +g_R \sigma_{1,3} +6 m_{H_u}^2 \mbox{Tr}\Big({Y_u  Y_{u}^{\dagger}}\Big) +2 m_{H_u}^2 \mbox{Tr}\Big({Y_v  Y_{v}^{\dagger}}\Big) +6 \mbox{Tr}\Big({T_u^*  T_{u}^{T}}\Big) +2 \mbox{Tr}\Big({T_\nu^*  T_{\nu}^{T}}\Big) \\ 
+2 \mbox{Tr}\Big({m_l^2  Y_{v}^{\dagger}  Y_v}\Big) +6 \mbox{Tr}\Big({m_q^2  Y_{u}^{\dagger}  Y_u}\Big) +6 \mbox{Tr}\Big({m_u^2  Y_u  Y_{u}^{\dagger}}\Big) +2 \mbox{Tr}\Big({m_{\nu}^2  Y_v  Y_{v}^{\dagger}}\Big)\,.
\end{multline}

\begin{multline}
\beta_{m_{\nu}^2}^{(1)} =  
-3 g_{BR}^{2} {\bf 1} |M_4|^2 +2 \sqrt{6} g_{BR} g_R {\bf 1} |M_4|^2 -2 g_{R}^{2} {\bf 1} |M_4|^2  
+\Big(-3 g_{BL}^{2} M_1  + g_{BL} \Big(2 \sqrt{6} g_{RB} M_1  -3 g_{BR} M_{B R}  + \sqrt{6} g_R M_{B R} \Big) \\ 
+ g_{RB} \Big(-2 g_{RB} M_1  -2 g_R M_{B R}  + \sqrt{6} g_{BR} M_{B R} \Big)\Big){\bf 1} M_1^* +\Big(-3 g_{BL}^{2} M_{B R} -3 g_{BR}^{2} M_{B R} 
+\sqrt{6} g_{BR} \Big(2 g_R M_{B R} \\ + g_{RB} \Big(M_1 + M_4\Big)\Big)-2 \Big(g_{R}^{2} M_{B R}  + g_{RB}^{2} M_{B R}  + g_R g_{RB} \Big(M_1 + M_4\Big)\Big) 
+g_{BL} \Big(-3 g_{BR} \Big(M_1 + M_4\Big) \\ + \sqrt{6} \Big(2 g_{RB} M_{B R}  + g_R \Big(M_1 + M_4\Big)\Big)\Big)\Big){\bf 1} M_{B R}^*  
-3 g_{BL} g_{BR} M_{B R} {\bf 1} M_4^* +\sqrt{6} g_{BL} g_R M_{B R} {\bf 1} M_4^*  \\ +\sqrt{6} g_{BR} g_{RB} M_{B R} {\bf 1} M_4^* -2 g_R g_{RB} M_{B R} {\bf 1} M_4^* +\sqrt{\frac{3}{2}} g_{BL} {\bf 1} \sigma_{1,1} - g_{RB} {\bf 1} \sigma_{1,1} +\sqrt{\frac{3}{2}} g_{BR} {\bf 1} \sigma_{1,3} - g_R {\bf 1} \sigma_{1,3} \\
+ 2 m_{\chi}^2 {Y_s  Y_{s}^{\dagger}}
+4 m_{H_u}^2 {Y_v  Y_{v}^{\dagger}} +2 {T_{s}  T_{s}^{\dagger}} +4 {T_\nu  T_{\nu}^{\dagger}} +{m_{\nu}^2  Y_s  Y_{s}^{\dagger}} +2 {m_{\nu}^2  Y_v  Y_{v}^{\dagger}} +2 {Y_s  m_{S}^2  Y_{s}^{\dagger}}   
+{Y_s  Y_{s}^{\dagger}  m_{\nu}^2} \\ +4 {Y_v  m_l^2  Y_{v}^{\dagger}} +2 {Y_v  Y_{v}^{\dagger}  m_{\nu}^2}\,.
\label{RGsneutrino}
\end{multline}

\begin{multline}
\beta_{m_l^2}^{(1)} =  
-3 g_{BR}^{2} {\bf 1} |M_4|^2 -6 g_{L}^{2} {\bf 1} |M_2|^2 -3 g_{BL} \Big(g_{BL} M_1  + g_{BR} M_{B R} \Big){\bf 1} M_1^* -3 \Big(g_{BL}^{2} M_{B R}  + g_{BL} g_{BR} \Big(M_1 + M_4\Big) + g_{BR}^{2} M_{B R} \Big){\bf 1} M_{B R}^* \\ 
-3 g_{BL} g_{BR} M_{B R} {\bf 1} M_4^* - \sqrt{\frac{3}{2}} g_{BL} {\bf 1} \sigma_{1,1} 
- \sqrt{\frac{3}{2}} g_{BR} {\bf 1} \sigma_{1,3}
+2 m_{H_d}^2 {Y_{e}^{\dagger}  Y_e} +2 m_{H_u}^2 {Y_{v}^{\dagger}  Y_v} +2 {T_{e}^{\dagger}  T_e} +2 {T_{\nu}^{\dagger}  T_\nu} +{m_l^2  Y_{e}^{\dagger}  Y_e} \\ 
+{m_l^2  Y_{v}^{\dagger}  Y_v}+2 {Y_{e}^{\dagger}  m_e^2  Y_e} +{Y_{e}^{\dagger}  Y_e  m_l^2}+2 {Y_{v}^{\dagger}  m_{\nu}^2  Y_v} +{Y_{v}^{\dagger}  Y_v  m_l^2} \,.
\label{RGslepton}
\end{multline}

\bibliography{mBLRISS}

\end{document}